\documentstyle[11pt]{article}

\def\be{\begin{equation}}
\def\ee{\end{equation}}
\renewcommand{\theequation}{\thesection.\arabic{equation}}
\newcommand{\bb}{\begin{equation}}
\newcommand{\bqn}{\begin{eqnarray}}
\newcommand{\eqn}{\end{eqnarray}}

\renewcommand{\theequation}{\thesection.\arabic{equation}}
\begin{document}
\begin{titlepage}

\begin{flushright}

HUTP-99/A050 \\
ULB-TH-99/14 \\
WIS/99--143

\end{flushright}

\begin{center}
{\Large {\bf Asymptotic Dynamics and Asymptotic Symmetries
of Three-Dimensional Extended AdS Supergravity}}

\end{center}
\vfill

\begin{center}
{\large
Marc Henneaux$^{a,b}$, Liat Maoz$^{c,d}$ and Adam Schwimmer$^{c}$
\footnote{henneaux@ulb.ac.be, maoz@wis.weizmann.ac.il,
maoz@physics.harvard.edu,
ftschwim@charm.weizmann.ac.il}}
\end{center}
\vfill

\begin{center}{\sl
$^a$ Physique Th\'eorique et Math\'ematique, Universit\'e Libre de
Bruxelles,\\
Campus Plaine C.P. 231, B--1050 Bruxelles, Belgium\\[1.5ex]

$^b$ Centro de Estudios Cient\'\i ficos de Santiago,\\
Casilla 16443, Santiago 9, Chile\\[1.5ex]

$^c$ Department of the Physics of Complex Systems,
Weizmann Institute, \\
Rehovot 76100, Israel\\[1.5ex]

$^d$ Jefferson Physical Laboratories, Harvard University \\
Cambridge, MA 02138, U.S.A.

}\end{center}

\vfill

\break

\begin{abstract}
We investigate systematically the asymptotic dynamics and symmetries of
all three-dimensional extended AdS supergravity models.
First, starting from the Chern-Simons formulation, we show
explicitly that the (super)anti-de Sitter boundary conditions imply
that the asymptotic symmetry algebra is the extended superconformal
algebra with quadratic nonlinearies in the currents.
We then derive the super-Liouville action by solving the
Chern-Simons theory and obtain a realization of the
superconformal algebras in terms of super-Liouville fields.
Finally, we discuss the possible periodic conditions that
can be imposed on the generators of the algebra and
generalize the spectral flow analysed previously in the context
of the $N$-extended linear superconformal algebras with $N \leq 4$.
The $(2+1)$-AdS/$2$-CFT correspondence
sheds a new light on the properties of the nonlinear
superconformal algebras.
It also provides a general and natural interpretation
of the spectral flow.

\end{abstract}
\vfill
\end{titlepage}

\section{Introduction}
\setcounter{equation}{0}

The connection between diffeomorphism-invariant theories in $D+1$
dimensions
and conformal field theories in $D$ dimensions is a fascinating subject
\cite{Maldacena,GP,Witten,review}.  One of the simplest contexts in which
it
has been studied is the case of $2+1$ dimensions, for which it has been
proved
more than ten years ago that pure Chern-Simons theory \cite{DJT} on
a manifold with boundary induces
the Wess-Zumino-Witten model \cite{W0} on the boundary
\cite{W1,MS,MSetAl}.
This context is also particularly rich since the conformal group on
the boundary is infinite-dimensional.

As it has been realized, the detailed correspondence between
$D+1$ and $D$ dimensions strongly depends on the precise form of the
boundary conditions.  So, while one gets the WZW model with the natural
boundary
conditions of \cite{MS,MSetAl}, one gets different theories if
one adopts other
boundary conditions.  This feature is relevant to $2+1$ gravity with
a negative cosmological constant (``AdS gravity in $2+1$ dimensions"),
which may be viewed as a Chern-Simons theory with gauge group
$SO(2,2)$ at least as far as the equations of motion are concerned.
\cite{AT,W2}.
It has been shown in \cite{CHvD} that the anti-de Sitter
asymptotics \cite{BH} lead to the Liouville theory at infinity (up to zero
modes).  In terms of the CS/WZW correspondence, one may undertand this
result
as due to the fact that the AdS asymptotic conditions are stronger
than the boundary conditions of \cite{MS,MSetAl} and enforce
the Hamiltonian reduction from $SL(2,R)$-WZW to Liouville theory
\cite{X,XX,XXX,XXXX}.

The purpose of this paper is to extend the analysis to the
supersymmetric case.   A first step in this direction was taken in
\cite{CH,BBCHO}, where the case $N=1$ was considered and shown to lead
either
to the Ramond or Neveu-Schwarz superalgebras at infinity,
depending on the periodicity conditions on the gravitino field.
The extended models were not treated in detail.  This is done here.

After a brief review of AdS supergravity (section 2),
we provide asymptotic conditions on all the fields.  These
conditions are geometrically motivated and chosen
to enforce AdS asymptotics.  We then show that they automatically imply
that the asymptotic superconformal algebras are the
superconformal algebras with non linearities
in the currents discussed in \cite{Y,YY,Frad1,Frad2,Frad3,Bowcock}
(section 3).
Furthermore, the (classical) Virasoro central charge is equal
to $6k= 3 \ell/2G$ for all models.
The analysis is done by identifying in the $(2+1)$-Chern-Simons
formulation the generators
of the asymptotic symmetries as appropriate surface terms
at infinity and then computing their algebra.
In sections 4 and 5, we give a dynamical explanation of the
origin of these
symmetries by relating explicitly the extended supergravity
actions to the extended
super-Liouville ones.  A few fine points (e.g., implementation
of the reduction constraints inside the action) are carefully analysed.
Finally, in section 6, we discuss the ``spectral flow" \cite{100}.
Different ``moddings" for the generators of the
superconformal algebras can be adopted, leading to apparently different
algebras.
However, some of these  algebras are in fact related by
the so-called spectral flow introduced in \cite{100} for the
linear superconformal
algebras.  We close our paper with a concluding section,
followed by an appendix, in which we discuss the role of 
zero modes and holonomies in the Liouville model arising from
$2+1$ gravity and their relation to the sign of the
Liouville potential.

The interest of extended AdS supergravity models in the context of the
AdS/CFT correspondence is at least threefold:
\begin{enumerate}
\item First, the treatment
sheds a new light on the properties of the nonlinear
algebras of \cite{Y,YY,Frad1,Frad2,Frad3,Bowcock}.
In particular, it provides a physical explanation for the relation
between the superconformal algebras and superalgebras found algebraically
in \cite{Frad2,Frad3,Bowcock}.
The superconformal algebra is the boundary symmetry of an AdS
supergravity theory in the Chern-Simons formulation with the superalgebra
as a gauge group.
\item Second, one automatically gets explicit realizations of these
superconformal algebras in terms of Liouville fields.
\item Third, the spectral flow  gets a general and natural interpretation
in the Chern-Simons framework.
\end{enumerate}

Some of the results reported here are implicitly
contained in the literature (see e.g. \cite{deBoer,Ishimoto,Ito,David}).
However,
no systematic and complete treatment is to our knowledge available.
In view of the recent interest in the AdS/CFT correspondence, we feel that
such a treatment can be useful.

\section{Extended AdS Supergravities in Three Dimensions}
\setcounter{equation}{0}

\subsection{Superalgebras}
As shown in \cite{AT}, supergravity in $2+1$ dimensions can be formulated
as a pure Chern-Simons theory based on an appropriate supergroup. The
dreibein, the gravitini and the gauge fields can be combined in a super
Chern-Simons connection.
We denote the superalgebra by
G=G$_{\overline{0}}\oplus$G$_{\overline{1}}$, where G$_{\overline{0}}$ is
the
even part of the superalgebra and G$_{\overline{1}}$ the odd part.
Since we
wish to describe AdS$_{3}$ space-time, the even subalgebra
G$_{\overline{0}}$ must contain $sl(2;R)$ and be of the form
G$_{\overline{0}}\cong$ $sl(2;R) \oplus\widetilde{G}$.  Furthermore,
the fermionic generators must transform as $sl(2;R)$-spinors.
The dimension of the ``internal" algebra
$\widetilde{G}$ is denoted $D=$ dim$\,\widetilde{G}$
and $\rho$ is the
real, not necessarily irreducible
representation of $\widetilde{G}$ (of dimension
$d=$ dim$\, \rho$) in which the fermionic generators transform.

These conditions are realized in only seven cases
\cite{Kac,Nahm,Gun}, which are the following

\vspace{.3cm}

\begin{tabular}
[c]{||c|c|c|c||}\hline\hline
G & $\widetilde{G}$ & $\rho$ & D\\\hline
Osp(N%
$\vert$%
2;R) & so(N) & N & N(N-1)/2\\\hline
SU(1,1%
$\vert$%
N)$_{N\neq2}$ & su(N)+u(1) & N+\={N} & N$^{2}$\\\hline
SU(1,1%
$\vert$%
2)/U(1) & su(2) & 2+\={2} & 3\\\hline
Osp(4$^{\ast}$%
$\vert$%
2M) & su(2)+usp(2M) & (2M,2) & M(2M+1)+3\\\hline
D$^{1}(2,1;\alpha)$ & su(2)+su(2) & (2,2) & 6\\\hline
G(3) & G$_{2}$ & 7 & 14\\\hline
F(4) & spin(7) & 8$_{s}$ & 21\\\hline\hline
\end{tabular}
\vspace{.3cm}

In fact, this is also the list of superalgebras which can be associated
with
2-dimensional superconformal algebras with quadratic non-linearities
\cite{Frad2,Frad3,Bowcock}.
This is not an accident since, as we shall see, the two are related
by the AdS/CFT correspondence.

\subsection{Conventions}
We follow \cite{Frad3} (see also \cite{BG}).
We denote the basis of the representation $\rho$ as
$\lambda^{a}=\left(  \lambda^{a}\right)  _{\beta}^{\alpha}$,
$\left[  \lambda^{a},\lambda^{b}\right]  =f^{abc}\lambda^{c}$, and take
the
structure constants $f^{abc}$ to be totally antisymmetric.
The Killing metric for $\widetilde{G}$ is
$g^{ab}=-f_{\quad}^{acd}f^{bcd}=-C_{v}\delta^{ab}$
where $C_{v}$ is the eigenvalue of the second Casimir in the adjoint
vrepresentation of $\widetilde{G}$.

We also denote by C$_{\rho}$ the eigenvalue of the second Casimir in the
representation $\rho$:
$$\lambda^{a}\lambda^{a}=-C_{\rho}I , \; \;
tr(\lambda^{a}\lambda^{b})=-\frac
{d}{D}C_{\rho}\delta^{ab}=\frac{d}{D}\frac{C_{\rho}}{C_{v}}g^{ab}.$$

The representation $\rho$ is orthogonal and admits
a $\widetilde{G}$-invariant symmetric metric
$\eta^{\alpha\beta}=\eta^{\beta\alpha}$, with inverse
$\eta_{\alpha\beta}$,
which we will use to lower and raise the supersymmetry (Greek) indices.
One has $\eta_{\alpha\beta}\eta^{\beta\gamma}=\delta_{\alpha}^{\gamma}$,
$\left(  \lambda^{a}\right)  ^{\alpha\beta}=\left(  \lambda^{a}\right)
_{\gamma}^{\alpha}\eta^{\gamma\beta}=-\left(  \lambda^{a}\right)
^{\beta\alpha}$, and also $\left(  \lambda^{a}\lambda^{b}\right)
^{\alpha\beta}=\left(  \lambda^{b}\lambda^{a}\right)  ^{\beta\alpha}$.

\vspace{.4cm}
The superalgebra generators and commutators are:

\begin{itemize}
\item $G_{\bar{0}}$ generators :
The sl(2;R) generators t$^{a}$ ($a=3,+,-$)
are equal to $\frac{1}{2}\sigma^{3},\sigma
^{+},\sigma^{-}$, and satisfy the commutation relations:
\begin{eqnarray}
\left[  \frac{1}{2}\sigma^{3},\sigma^{\pm}\right]   =\pm\sigma^{\pm}\\
\left[  \sigma^{+},\sigma^{-}\right]    =2\left(  \frac{1}{2}\sigma
^{3}\right)
\end{eqnarray}
The Killing metric for sl(2;R) is:$$h^{ab}=\left(
\begin{array}
[c]{ccc}%
2 & 0 & 0\\
0 & 0 & 4\\
0 & 4 & 0
\end{array}
\right), $$
and in its 2-representation, Tr$\left(  t^{a}t^{b}\right)  =\frac
{1}{4}h^{ab}$

The $\widetilde{G}$ generators are denoted $T^{a}$
($a=1...D$), and
 satisfy:
\begin{equation}
\left[  T^{a},T^{b}\right]  =f^{abc}T^{c} \label{com1}
\end{equation}
These commute with all sl(2;R) generators.

\item $G_{\bar{1}}$ generators :
These are denoted by $R^{\pm\alpha}$
($\alpha=1...d$).
Their commutators with the sl(2;R) generators are:%
\begin{eqnarray}
\left[  \frac{1}{2}\sigma^{3},R^{\pm\alpha}\right]    &=&\pm\frac{1}
{2}R^{\pm\alpha}\label{com2}\\
\left[  \sigma^{\pm},R^{\pm\alpha}\right]    &=& 0\\
\left[  \sigma^{\pm},R^{\mp\alpha}\right]    &=& R^{\pm\alpha}.
\end{eqnarray}
Their commutators with the $\widetilde{G}$ generators are:
\begin{equation}
\left[  T^{a},R^{\pm\alpha}\right]  =-\left(  \lambda^{a}\right)  _{\beta
}^{\alpha}R^{\pm\beta} \label{com3}
\end{equation}
The anticommutators of the fermion generators are:
\begin{eqnarray}
\left\{  R^{\pm\alpha},R^{\pm\beta}\right\}   & =& \pm\eta^{\alpha\beta}
\sigma^{\pm}\label{com4}\\
\left\{  R^{\pm\alpha},R^{\mp\beta}\right\}    &=&
-\eta^{\alpha\beta}\left(
\frac{1}{2}\sigma^{3}\right)  \pm\frac{d-1}{2C_{\rho}}\left(  \lambda
^{a}\right)  ^{\alpha\beta}T^{a}
\end{eqnarray}
The Jacobi identity for three fermion generators yields an identity
involving
the representation matrices, and is:
\begin{equation}
\left(  \lambda^{a}\right)  ^{\beta\gamma}\left(  \lambda^{a}\right)
_{\delta}^{\alpha}+\left(  \lambda^{a}\right)  ^{\alpha\gamma}\left(
\lambda^{a}\right)  _{\delta}^{\beta}=\frac{C_{\rho}}{d-1}\left(
2\eta^{\alpha\beta}\delta_{\delta}^{\gamma}-\eta^{\alpha\gamma}\delta_{\delta
}^{\beta}-\eta^{\beta\gamma}\delta_{\delta}^{\alpha}\right)
\label{Jacobi}%
\end{equation}
It is fulfilled for all supergroups listed above.

\end{itemize}

The given superalgebras all admit a consistent, invariant, supersymmetric
and nondegenerate bilinear form which we denote by $STr$ and which is
defined by
\begin{equation}
STr\left(  t^{a}t^{b}\right)  =\frac{1}{4}h^{ab}\quad;\quad STr(R^{-\alpha
}R^{+\beta})=-STr(R^{+\alpha}R^{-\beta})=\eta^{\alpha\beta}\quad;\quad
STr(T^{a}T^{b})=\frac{2C_{\rho}}{d-1}\delta^{ab} \label{form}%
\end{equation}

\subsection{Action}
Using these conventions, we can parametrize a general element in G by:
\begin{equation}
\Gamma=\left(  A^{3}\frac{\sigma^{3}}{2}+A^{+}\sigma^{+}+A^{-}\sigma
^{-}\right)  +\left(  B^{a}T^{a}\right)  +\left(  \psi_{+\alpha}R^{+\alpha
}+\psi_{-\alpha}R^{-\alpha}\right)  \equiv A+B+\Psi\label{comps}%
\end{equation}
where $A\equiv A^{3}\frac{\sigma^{3}}{2}+A^{+}\sigma^{+}+A^{-}\sigma^{-}%
\quad,\quad B\equiv
B^{a}T^{a}\quad,\quad\Psi\equiv\psi_{+\alpha}R^{+\alpha
}+\psi_{-\alpha}R^{-\alpha}.$
Here \bigskip$A^{3},A^{+},A^{-},B^{a}$ are commuting parameters and
$\psi_{+\alpha},\psi_{-\alpha}$ anti-commuting Grassman parameters.

The action describing supergravity in AdS$_{3}$ space-time is a
difference of two Chern-Simons actions \cite{AT},
\begin{equation}
S[\Gamma,\widetilde{\Gamma}]=S_{CS}[\Gamma]-S_{CS}[\widetilde{\Gamma}]
\label{diff}%
\end{equation}
where $\Gamma,\widetilde{\Gamma}\in G$
are superalgebra valued super-connections.
We assume for simplicity that $\Gamma$ and $\widetilde{\Gamma}$
take values in the same superalgebra, although this is not necessary.
This choice leads to a non-chiral model in two dimensions
to which the results of sections 4 and 5 apply.
$S_{CS}$ is the super Chern-Simons action, defined by:%
\begin{equation}
S_{CS}[\Gamma]=\frac{k}{4\pi}{\int}_{M}STr\left(  \Gamma\wedge
d\Gamma+\frac{2}{3}\Gamma\wedge\Gamma\wedge\Gamma\right)  \label{scsact}%
\end{equation}
The integration is over M - a 3-manifold, with topology $D\times\Re$
and $k$ is related to the $3$-dimensional Newton constant $G$
through $k = \ell/4G$ where $\ell$ is the anti-de Sitter radius.
Furthermore, the product of two fermions in this
formula (as well as in (\ref{SSSS}) below but in no
other formula) differs by a factor $i$ from the standard Grassmann product
fulfilling $(ab)^* = b^* a^*$
(e.g., (\ref{scsact}) contains $i (k/4\pi)  \psi_{+ \alpha}
\wedge d \psi_{-\beta} \; STr(R^{+ \alpha} R^{-\beta})$ in terms
of the standard Grassmann product).

Writing the super-connection in component form
and identifying the $sl_{2}$ components with the dreibeins and the
connections:
\begin{eqnarray}
A_{i}^{a}  & = & \omega_{i}^{a}+\frac{1}{\ell}e_{i}^{a}
\label{dreib} \\
\widetilde{A}_{i}^{a}  & = & \omega_{i}^{a}-\frac{1}{\ell}e_{i}^{a},
\label{dreibein}
\end{eqnarray}
we find the supergravity action for a general superalgebra:
\begin{eqnarray}
S\left[  \Gamma,\widetilde{\Gamma}\right]   & = & \frac{1}{8\pi G}
{\int}_{M}d^{3}x\{\frac{1}{2}eR+\frac{e}{\ell^{2}}+\nonumber\\
& & \quad\ \ \ \ \ \ \ \ \ \ \ \ \ \ \ \ \ \ \ \ -\frac{i}{2}\ell
\varepsilon^{ijk}\left(  \psi_{i}\right)  _{\mu}{\cal
D}_{j}^{\mu\nu}\left(
\psi_{k}\right)  _{\nu}+\frac{i}{2}\ell\varepsilon^{ijk}\left(
\widetilde{\psi}_{i}\right)  _{\mu}\widetilde{{\cal D}}_{j}^{\mu\nu}\left(
\widetilde{\psi}_{k}\right)_{\nu} \nonumber\\
&  &\quad\ \ \ \ \ \ \ \ \ \ \ \ \ \ \ \ \ \ \ \ +\frac{C_{\rho}}{d-1}
\ell\varepsilon^{ijk}\left(  B_{i}^{a}\partial_{j}B_{k}^{a}+\frac{1}
{3}f^{abc}B_{i}^{a}B_{j}^{b}B_{k}^{c}\right) \nonumber\\
&  &\;\ \;\ \ \ \ \ \ \ \ \ \ \ \ \ \ \ \ \ \ \ -\frac{C_{\rho}}{d-1}
\ell\varepsilon^{ijk}\left(  \widetilde{B}_{i}^{a}\partial_{j}\widetilde
{B}_{k}^{a}+\frac{1}{3}f^{abc}\widetilde{B}_{i}^{a}\widetilde{B}_{j}
^{b}\widetilde{B}_{k}^{c}\right) \nonumber\\
&  &\;\ \;\ \ \ \ \ \ \ \ \ \ \ \ \ \ \ \ \ \ \ -\frac{i}{2}\varepsilon
^{ijk}\eta^{\alpha\beta}e_{i}^{a}\left(  \left[  \bar{\psi}_{j}\right]
_{\alpha}t^{a}\left[  \psi_{k}\right]  _{\beta}-\left[  \overline
{\widetilde{\psi}}_{j}\right]  _{\alpha}t^{a}\left[  \tilde{\psi}
_{k}\right]  _{\beta}\right)  \} \label{ActionAction}
\end{eqnarray}
where the square brackets stand for two-component $sl_2$-spinors and
where the operators ${\cal D}_{j}^{\mu\nu}$ and
$\widetilde{{\cal D}}_{j}^{\mu\nu}$ are respectively defined by
\begin{equation}
{\cal D}_{j}^{\mu\nu}\equiv\left[
\begin{array}
[c]{c}%
2\left(  \eta^{\alpha\beta}\partial_{j}+\left(  \lambda^{a}\right)
^{\alpha\beta}B_{j}^{a}\right)
\delta_{+\alpha}^{\mu}\delta_{-\beta}^{\nu}+\\
-\eta^{\alpha\beta}\left(  \frac{1}{2}\omega_{j}^{3}\left[
\delta_{+\alpha
}^{\mu}\delta_{-\beta}^{\nu}+\delta_{-\alpha}^{\mu}\delta_{+\beta}^{\nu
}\right]  +\omega_{j}^{+}\delta_{-\alpha}^{\mu}\delta_{-\beta}^{\nu}%
-\omega_{j}^{-}\delta_{+\alpha}^{\mu}\delta_{+\beta}^{\nu}\right)
\end{array}
\right]
\end{equation}%
\begin{equation}
\widetilde{{\cal D}}_{j}^{\mu\nu}\equiv\left[
\begin{array}
[c]{c}%
2\left(  \eta^{\alpha\beta}\partial_{j}+\left(  \lambda^{a}\right)
^{\alpha\beta}\widetilde{B}_{j}^{a}\right)  \delta_{+\alpha}^{\mu}%
\delta_{-\beta}^{\nu}+\\
-\eta^{\alpha\beta}\left(  \frac{1}{2}\omega_{j}^{3}\left[
\delta_{+\alpha
}^{\mu}\delta_{-\beta}^{\nu}+\delta_{-\alpha}^{\mu}\delta_{+\beta}^{\nu
}\right]  +\omega_{j}^{+}\delta_{-\alpha}^{\mu}\delta_{-\beta}^{\nu}%
-\omega_{j}^{-}\delta_{+\alpha}^{\mu}\delta_{+\beta}^{\nu}\right)
\end{array}
\right]
\end{equation}

(\bigskip From here on we work in units of $\ell=1$, unless otherwise
stated).

\section{Asymptotic Symmetries - Extended superconformal algebras}
\setcounter{equation}{0}

\subsection{Boundary conditions}
The ground state of AdS supergravity is anti-de Sitter space with
vanishing gravitini and gauge fields.  AdS$_3$ is the
solution with the maximum number of isometries, namely $3+3 = 6$
($3$ per chiral sector).
It is also invariant under constant $\widetilde{G}
\oplus \widetilde{G}$-transformations
($D+D$ tansformations) as well as
under $2d + 2d$ rigid supersymmetries.
The corresponding Killing spinors are explicitly given in
\cite{CH}.  In short, AdS$_3$ is invariant under the superalgebra
$G \oplus G$.

Besides AdS$_3$,
there are other solutions,
which differ from AdS$_3$ in their global properties
\cite{DJ}.  The black-hole solution \cite{BTZ}
is an important example and can be obtained from AdS$_3$ through
appropriate identifications \cite{BTZH}.
The boundary conditions to be imposed on the fields at infinity should
allow for these physically interesting solutions.

It was shown in \cite{BH} that the appropriate conditions in the
$sl(2;R)$-sector are, for $r\rightarrow \infty$,
\begin{eqnarray}
A &\sim &
\left\{  \frac{1}{r} \frac{2 \pi L(\theta,t)}{k}
\sigma^{+}+r\sigma^{-} \right\}dx^{+}
+0dx^{-}
+\left\{  \frac{1}{r}\frac{\sigma^{3}}{2}\right\}  dr
\\
\tilde{A}
& \sim &
\left\{  \frac{1}{r}\frac{ 2 \pi \widetilde
{L}(\theta,t)}{k}\sigma^{-}+r\sigma^{+} \right\} dx^{-}
+0dx^{+}
+\left\{ -\frac{1}{r}\frac{\sigma^{3}}{2}\right\}  dr
\end{eqnarray}
where we have switched to chiral coordinates: $x^{\pm}=t\pm\ell\theta$
and where $L$ and $\widetilde{L}$ are arbitrary functions of $\theta$
and $t$.  The factors of $2 \pi$ are introduced for later
convenience.
(The asymptotic conditions were actually expressed in terms of the
metric in \cite{BH} but are easily verified to be equivalent
to those written here in terms of the connection).

In order to obtain the asymptotic boundary conditions for all components
of
the two superconnections, we follow the general scheme introduced in
\cite{Teit,BBCHO}.  Namely, we take a generic configuration having
the above asymptotic behaviour with vanishing
gravitini and gauge fields
and act on it with the exact symmetries of AdS$_3$.
In doing so, one generates
new terms (since we start with configurations which are not AdS$_3$)
which typically behave as
\begin{eqnarray}
& & \Gamma\sim\left\{  \frac{1}{r} \frac{2 \pi L(\theta,t)}{k}
\sigma^{+}+r\sigma^{-}
+\frac{1}{\sqrt{r}}\frac{2 \pi Q_{+\alpha}\left(
\theta,t\right)}{k}R^{+\alpha}
+\frac{2 \pi B^{a}\left(  \theta,t\right)  }{k_{B}}T^{a}\right\}dx^{+}
\nonumber \\
& & \quad+0dx^{-}\nonumber\\
& & \quad+\left\{  \frac{1}{r}\frac{\sigma^{3}}{2}\right\}  dr
\label{asymp}
\end{eqnarray}
\begin{eqnarray}
& & \widetilde{\Gamma}\sim\left\{  \frac{1}{r}\frac{2 \pi \widetilde
{L}(\theta,t)}{k}\sigma^{-}+r\sigma^{+}+\frac{1}{\sqrt{r}}\frac
{2 \pi \widetilde{Q}_{-\alpha}\left(  \theta,t\right)  }{k}R^{-\alpha}%
+\frac{2 \pi \widetilde{B}^{a}\left(  \theta,t\right)
}{k_{B}}T^{a}\right\}
dx^{-} \nonumber \\
& & \quad+0dx^{+}\quad\nonumber\\
& & \quad+\left\{  -\frac{1}{r}\frac{\sigma^{3}}{2}\right\}  dr
\label{asymp2}
\end{eqnarray}
where $L(\theta,t)$, $Q_{+\alpha}\left(
\theta,t\right)$, $B^{a}\left(  \theta,t\right)$
and $\widetilde{L}(\theta,t)$, $\widetilde{Q}_{-\alpha}\left(
\theta,t\right)$, $\widetilde{B}^{a}\left(  \theta,t\right)$
are functions of $\theta,t$ ($Q,\widetilde{Q}$ Grassmanian), so
we shall adopt (\ref{asymp}) and (\ref{asymp2}) with arbitrary $L$,
$\widetilde{L}$, $Q$, $\widetilde{Q}$, $B$ and $\widetilde{B}$
as boundary conditions.  Here, we have scaled the terms proportional
to $B^a$ and $\widetilde{B}^a$ with
\begin{equation}
k_B= k \frac{2C_\rho}{d-1}
\end{equation}
for
later convenience.

\subsection{Asymptotic Symmetries}

The boundary conditions at spatial infinity give us the asymptotic
form of the superconnections. We work out in this subsection the
gauge transformations that preserve the asymptotic form of the
superconnections (``asymptotic symmetries").
In what follows, we focus only on
the superconnection $\Gamma$. The treatment for $\tilde{\Gamma}$ is
analogous
and eventually gives another copy of the superconformal
algebra for the other chirality.

It is convenient to factor out the $r$ dependence by performing a gauge
transformation and to work with the equivalent
connection $\Delta_i$ defined by
\begin{equation}
\Delta_i = b \partial_i b^{-1} + b \Gamma_i b^{-1}
\end{equation}
where b is a similarity transformation depending only on r:%
\begin{equation}
b(r)=\exp\left[  \frac{\sigma^{3}}{2}\ln r\right].
\end{equation}
The components of $\Delta_i$ vanish asymptotically, except
$\Delta_u$ ($u \equiv x^+$), given by
\begin{equation}
\frac{1}{2\pi}\Delta_{u}=\frac{1}{2\pi}b\Gamma_{u}b^{-1}
= \left(  \frac{L}{k}\sigma^{+}+\frac{1}{2\pi}\sigma^{-}\right)
+\left(  \frac{B^{a}}{k_{B}}T^{a}\right)  +\left(
\frac{Q_{+\alpha}}{k}R^{+\alpha}\right) .
\end{equation}
Two things should be mentioned about the asymptotic form of
$\Delta_u$.  First, there is no $\sigma^3$ component and the term
proportional to $\sigma^-$ is fixed to be one.  Second, the fermion
part of the superconnection
contains no $R^{-\alpha}$ components. These
conditions turn out to be crucial for obtaining the superconformal
algebra.
Note that there is a similar ``chirality condition" on the fermions
in $4$ (or higher) dimensions \cite{Teit}.

Acting on the superconnection with an infinitesimal gauge transformation:
\begin{equation}
\Lambda=\left(  \chi^{3}\frac{\sigma^{3}}{2}+\chi^{+}\sigma^{+}+\chi
^{-}\sigma^{-}\right)  +\left(  \omega^{a}T^{a}\right)  +\left(
\epsilon_{+\alpha}R^{+\alpha}+\epsilon_{-\alpha}R^{-\alpha}\right)
\end{equation}
we get:
\begin{eqnarray}
\delta\Delta_{u} & =&\partial_u\Lambda+\left[
\Delta_{u},\Lambda\right]   \label{SSSS}\\
&  =&\left(  \partial\chi^{3}+2\frac{2\pi L}{k}\chi^{-}-2\chi^{+}-i\eta
^{\alpha\beta}\frac{2\pi Q_{+\alpha}}{k}\epsilon_{-\beta}\right)  \frac{1}
{2}\sigma^{3}\nonumber\\
&  +&\left(  \partial\chi^{+}-\frac{2\pi L}{k}\chi^{3}+i\eta^{\alpha\beta
}\frac{2\pi Q_{+\alpha}}{k}\epsilon_{+\beta}\right)  \sigma^{+}\nonumber\\
&  +&\left(  \partial\chi^{-}+\chi^{3}\right)  \sigma^{-}\nonumber\\
&  +&\left(  \partial\omega_{c}+f_{\quad}^{abc}\frac{2\pi
B^{a}}{k_B}\omega
^{b}+i\frac{d-1}{2C_{\rho}}\left(  \lambda^{c}\right)  ^{\alpha\beta}
\frac{2\pi Q_{+\alpha}}{k}\epsilon_{-\beta}\right)  T^{c}\nonumber\\
&  +&\left(  \partial\epsilon_{+\beta}+\frac{2\pi L}{k}\epsilon_{-\beta
}-\left(  \lambda^{a}\right)  _{\beta}^{\alpha}\frac{\epsilon_{+\alpha}
}{k_B}2\pi B^{a}-\chi^3\frac{2\pi Q_{+\beta}}{2k}+\left(
\lambda^{a}\right)
_{\beta}^{\alpha}\omega^{a}\frac{2\pi Q_{+\alpha}}{k}\right)  R^{+\beta
}\nonumber\\
&  +&\left(  \partial\epsilon_{-\beta}+\epsilon_{+\beta}-\left(  \lambda
^{a}\right)  _{\beta}^{\alpha}\epsilon_{-\alpha}\frac{2\pi
B^{a}}{k_{B}}-\chi
^{-}\frac{2\pi Q_{+\beta}}{k}\right)  R^{-\beta}
\end{eqnarray}
In order to preserve the asymptotic form of the superconnection,
the parameters of the gauge
transformation $\Lambda$ are required to fulfill some relations.
These relations are easily displayed by
taking as independent parameters
$\chi\equiv\chi^{-}$, $\epsilon_{\alpha}\equiv\epsilon_{-\alpha}$,
and $\omega^{a}$.  The relations in question are then
just expressions for the other parameters,
\begin{eqnarray}
\chi^{3}  &  = &-\chi^{\prime}\label{subgrp}\\
\chi^{+}  &  = &-\frac{\chi^{\prime\prime}}{2}+\chi\frac{2\pi L}{k}%
-i\eta^{\alpha\beta}\frac{2\pi Q_{+\alpha}}{2k}\epsilon_{\beta}
\label{subgrp1}\\
\epsilon_{+\alpha} &  = &-\epsilon_{\alpha}^{\prime}+\chi\frac{2\pi
Q_{+\alpha}
}{k}+\left(  \lambda^{a}\right)  _{\alpha}^{\beta}\epsilon_{\beta}
\frac{2\pi B_{a}}{k_{B}} \label{subgrp2}
\end{eqnarray}
where the prime denotes derivative with respect to the argument.
The transformations with parameters subject to (\ref{subgrp}),
(\ref{subgrp1}) and (\ref{subgrp2}) are the asymptotic symmetries.

\subsection{Superconformal Algebra}
In order to exhibit the algebra of the asymptotic symmetry
transformations,
we shall work out the Poisson bracket algebra of their generators.
To that end, we first need to identify these generators.  This
is done by following the Regge-Teitelboim method \cite{RT}.  The
gauge transformations of the Chern-Simons theory with parameters
$\Lambda^A \equiv ( \chi^3, \chi^{\pm}, \omega^a,
\epsilon_{\pm \alpha} )$ are generated in the
equal-time Poisson bracket by the spatial integral
$G[\Lambda] = \int d^2 x \Lambda^A {\cal G}_A + S_{\infty}$, where
(i) the ${\cal G}_A$ are the Chern-Simons constraints, equal to
minus the factor of the temporal
components of the superconnection in the action; and (ii) $S_{\infty}$
is a boundary term at infinity
chosen such that the variation $\delta G[\Lambda]$
of the generator $G[\Lambda]$ contains only undifferentiated field
variations under the given boundary conditions \cite{RT} (``$G[\Lambda]$
has well-defined functional derivatives").  If one follows this procedure,
one
gets (up to the bulk term that vanishes on-shell),
\begin{eqnarray}
G_{L}\left[  \chi\right] &  =&\int d\theta \;
\chi(\theta) L(\theta) \label{gen1}\\
G_{B}\left[  \omega\right] &  =&\int d\theta \;
\omega^{a}(\theta)B^{a}(\theta)\label{gen2}\\
G_{Q}\left[  \epsilon\right] &  =&\int d\theta \;
\eta^{\alpha\beta}i\epsilon_{\alpha}(\theta)Q_{+\beta}(\theta).
\label{gen3}
\end{eqnarray}
Thus, we see that $L$, $Q_{+ \alpha}$ and $B^a$ are precisely
the generators of the asymptotic symmetries.  There is no factor
in (\ref{gen1}), (\ref{gen2}) or (\ref{gen3}) because we
included the appropriate factors in the definitions
of $L$, $Q_{+ \alpha}$ and $B^a$.

Once the generators have been identified, one computes their
Poisson bracket algebra by using the relationship
$\delta_\Lambda F = \{G[\Lambda], F\}_{PB}$ , valid for any
phase-space function, and working out $\delta L$,
$\delta Q_{+ \alpha}$ and $\delta B^a$ directly
from the formulas of the previous subsections.

Under the asymptotic AdS subgroup, the transformation
laws for the L, Q's and B's are given by:
\begin{eqnarray}\
\delta L  &  =& - \frac{k}{2\pi} \frac{\chi'''}{2} + \left[ \left( \chi L
\right) ' +\chi 'L \right] -i\eta^{\alpha\beta} \left[
\frac{1}{2} \left( Q_{+\alpha}\epsilon_{\beta} \right) '+
Q_{+\alpha}\epsilon_{\beta}^{\prime} \right] \nonumber\\ &  -&
\frac{2\pi i}{k_{B}}\left( \lambda^{a}\right)
^{\alpha\beta} Q_{+\alpha} B^{a}  \epsilon_{\beta}\\
\delta Q_{+\alpha}  &  =& -\frac{k}{2\pi}\epsilon_{\alpha}^{\prime\prime}
+\left[  \left( \chi Q_{+\alpha}\right)
^{\prime}+\frac{1}{2}\chi^{\prime}Q_
{+\alpha}\right] +L\epsilon_{\alpha}+\nonumber\\
&  +& \left(  \lambda^{a}\right)  _{\alpha}^{\beta}\frac{k}{k_B}\left[
\left(  \epsilon_{\beta}B^{a}\right)  ^{\prime}+\epsilon_{\beta}^{\prime}
B^{a}\right]  +\nonumber\\
&  +& \left(  \lambda^{a}\right)  _{\alpha}^{\beta}\omega^{a}Q_{+\beta}%
-\frac{2\pi}{k_{B}}\chi\left(  \lambda^{a}\right)  _{\alpha}^{\beta}
Q_{+\beta} B^{a}  -2\pi \frac{k}{2k_{B}^{2}}\left\{  \lambda
^{a},\lambda^{b}\right\}  _{\alpha}^{\gamma}\epsilon_{\gamma}B^{a}B^{b}\\
\delta B^{a}  &  =& \frac{k_B}{2\pi}
\omega^{a\prime}+f^{abc}B^{b}\omega^{c}+i\frac{k_B}{k}
\frac{d-1}{2C_{\rho}}\left( \lambda^{a}\right)  ^{\alpha\beta}
Q_{+\alpha}\epsilon_{\beta}
\end{eqnarray}
The formula $\delta_\Lambda F = \{G[\Lambda], F\}_{PB}$ gives
then
\begin{eqnarray}
\left\{ L(\theta),L(\theta')\right\} _{PB}
&=& \frac{k}{4\pi}\delta'''\left( \theta-\theta'\right)
-\left(L(\theta)+L(\theta')\right) \delta'(\theta-\theta')\\
i\left\{ Q_{+\alpha}(\theta),Q_{+\beta}\left( \theta'\right) \right\}
_{PB}
 &  =&
-\frac{k}{2\pi}\eta_{\alpha\beta}\delta''(\theta-\theta')
-\left( \lambda^{a}\right)_{\alpha\beta}\frac{k}{k_B}
\delta'(\theta-\theta')\left[ B^{a}(\theta)+B^a(\theta')
\right]\nonumber\\
&  +& \eta_{\alpha\beta}L(\theta)\delta(\theta-\theta')
-2\pi\frac{k}{2k_{B}^2}\left\{ \lambda^{a},\lambda^{b}\right\}
_{\alpha\beta}B^{a}(\theta)B^{b}(\theta)\delta(\theta-\theta')\nonumber\\
\\
\left\{ B^{a}( \theta),B^{b}( \theta') \right\} _{PB} &=&
-\frac{k_B}{2\pi}
\delta^{ab}\delta'(\theta-\theta')+f^{abc}\delta(\theta-\theta')B^c(\theta)\\
\left\{ L( \theta),Q_{+\alpha}(\theta')\right\} _{PB} &  =&
-\left[ Q_{+\alpha}\left( \theta\right) +\frac{1}{2} Q_{+\alpha}
\left( \theta' \right) \right] \delta'\left( \theta-\theta' \right)
\nonumber\\
&  -& 2\pi\left( \lambda^{a}\right) _{\alpha}^{\beta}\frac{1}{k_B}
Q_{+\beta}(\theta) B^a(\theta) \delta \left( \theta-\theta' \right) \\
\left\{ B^a(\theta),Q_{+\alpha}(\theta') \right\} _{PB}
&=& \left( \lambda^{a}\right)  _{\alpha}^{\beta}Q_{+\beta}(\theta)
\delta(\theta-\theta')\\
\left\{ L\left( \theta\right) ,B^a\left( \theta'\right) \right\} _{PB}
&=& 0
\end{eqnarray}

As defined, the generator $L$ does not act on the Kac-Moody currents
$B^a$.
To get exactly the superconformal algebra in
standard form, we must add to $L$ the Sugawara energy-momentum
operator for the $B^a$, given classically by
$L_{SUG}=\frac{2\pi}{2k_B}B^{a}B^{a}$.
Thus, redefining $L$ by adding to it $L_{SUG}$:
\begin{equation}
\widehat{L}\equiv L+L_{SUG}=L+\frac{2\pi}{2k_B}B^{a}B^{a}
\end{equation}
and rescaling $Q$: $\widehat{Q}_{+\alpha}=\sqrt{2}Q_{+\alpha}$,
the Poisson bracket algebra becomes
\begin{eqnarray}
\left\{ \widehat{L}(\theta),\widehat{L}(\theta')\right\} _{PB}
& =& \frac{k}{4\pi}\delta'''(\theta-\theta')
-\left( \widehat{L}(\theta)+\widehat{L}(\theta')\right)
\delta'(\theta-\theta')\\
i\left\{ \widehat{Q}_{+\alpha}(\theta),\widehat{Q}_{+\beta}(\theta')
 \right\} _{PB}&  =&
-\frac{k}{\pi}\eta_{\alpha\beta}\delta''(\theta-\theta')
-2\left(  \lambda^{a}\right)_{\alpha\beta}\frac{d-1}{2C_{\rho}}
\delta'(\theta-\theta')\left[ B^{a}(\theta)+B^a(\theta')
\right]\nonumber\\
&  +&
2\eta_{\alpha\beta}\widehat{L}(\theta)\delta(\theta-\theta')\nonumber\\
&  -& 2\pi k  \left( \frac{d-1}{2kC_{\rho}} \right) ^2 \left[
\left\{ \lambda^{a},\lambda^{b} \right\}
_{\alpha\beta} + \frac{2C_{\rho}}{d-1} \eta_{\alpha\beta}\delta^{ab}
\right]
\nonumber \\ &{}& \;\; \; \; \; \; \; \; \times
B^{a}(\theta)B^{b}(\theta)\delta(\theta-\theta')\\
\left\{ B^{a}\left( \theta\right),B^{b}( \theta')\right\} _{PB}
& =& -\frac{k}{2\pi}\frac{2C_{\rho}}{d-1}
\delta^{ab}\delta'(\theta-\theta')
+f^{abc}\delta(\theta-\theta')B^c(\theta)\\
\left\{ \widehat{L}\left( \theta\right)
,\widehat{Q}_{+\alpha}\left( \theta'\right) \right\} _{PB}
&  =& -\left[ \widehat{Q}_{+\alpha}\left( \theta\right)
+\frac{1}{2} \widehat{Q}_{+\alpha} \left( \theta' \right) \right]
\delta'\left( \theta-\theta' \right) \\
\left\{ B^{a}\left( \theta\right) ,\widehat{Q}_{+\alpha}\left(
\theta'\right)
\right\} _{PB}  &=& \left(  \lambda^{a}\right)  _{\alpha}^{\beta}
\widehat{Q}_{+\beta}(\theta)\delta(\theta-\theta')\\
\left\{ \widehat{L} ( \theta),B^a(\theta') \right\} _{PB}  &=&
-B^a(\theta)\delta'(\theta-\theta')
\end{eqnarray}

The Fourier mode form of this algebra is given,
in quantum-mechanical notations, by\footnote{We set
$A(\theta)=\frac{1}{2\pi}\sum_{n}A_ne^{in\theta}$
for any operator $A(\theta)$ and use the
correspondence $\left\{ ,\right\} _{PB}=-i[,]$
(where $[,]$ is the commutator) for any pair of operators, except
when both are fermionic in which case one has
$\left\{ ,\right\} _{PB}=-i\left\{ ,\right\} $
(anticommutator).}
\begin{eqnarray}
\lbrack \widehat{L}_{m},\widehat{L}_{n}]  &  =&(m-n)\widehat{L}
_{m+n}+\frac{k}{2}m^3\delta_{m+n,0}\\
\left\{  \left( \widehat{Q}_{+\alpha}\right)  _{m},\left( \widehat
{Q}_{+\beta}\right)  _{n}\right\}   &  =&2\eta_{\alpha\beta}\widehat{L}
_{m+n}-2i\frac{d-1}{2C_{\rho}}(m-n)\left(  \lambda^{a}\right)
_{\alpha\beta
}\left(  B^{a}\right)  _{m+n}\nonumber\\
&  +&2k\eta_{\alpha\beta}m^2\delta_{m+n,0}  +\nonumber\\
& -& k \left( \frac{d-1}{2kC_{\rho}}\right) ^2 \left[
\left\{ \lambda^{a},\lambda^{b} \right\}
_{\alpha\beta} + \frac{2C_{\rho}}{d-1} \eta_{\alpha\beta}\delta^{ab}
\right]
\left(  B^{a}B^{b}\right)  _{m+n}\nonumber\\
\left[    B^{a} _{m},  B^{b}  _{n}\right]   &
=& i f^{abc} B^{c}_{m+n}
+\frac{2C_{\rho}k}{d-1}m\delta^{ab}\delta_{m+n,0}\\
\left[ \widehat{L}_{m},\left( \widehat{Q}_{+\alpha}\right)
_{n}\right]   &  =&(\frac{m}{2}-n)\left( \widehat{Q}_{+\alpha}\right)
_{m+n}\\
\left[ B^{a}_{m},\left( \widehat{Q}_{+\alpha}\right)
_{n}\right]   &  =&i \left(  \lambda^{a}\right)  _{\alpha}^{\beta}\left(
\widehat{Q}_{+\beta}\right)  _{m+n}=- i \eta_{\gamma\alpha}\left(
\lambda
^{a}\right)  _{\beta}^{\gamma}\left( \widehat{Q}^{+\beta}\right)
_{m+n}\\
\left[ \widehat{L}_{m}, B^{a}_{n}\right]   &  =&-n
B^{a}_{m+n}
\end{eqnarray}
Here, by $\left(  B^{a}B^{b}\right)  _{m+n}$,
we mean $B^{a}(\theta)B^{b}(\theta)=\left(
\frac{1}{2\pi} \right) ^2 \sum_n \left( B^aB^b \right)_n
e^{in\theta}.$
These are the non-linear superconformal algebras of
\cite{Y,YY,Frad2,Frad3,Bowcock}, up to a constant shift of
$\widehat{L}_0$.  The Virasoro central
charge is equal to $6k$ for all models.
It should be stressed, however, that
we obtained the classical version of the algebra, where
quantum effects of normal ordering are missing.
Such effects modify the value
of the coefficients in the algebra.

Thus, we have established that the boundary conditions of asymptotic AdS
on
$\Gamma$ lead to an asymptotically superconformal symmetry algebra of the
type discussed in \cite{Y,YY,Frad1,Frad2} and with classical Virasoro
central
charge equal to $6k$.
Repeating this treatment
for $\tilde{\Gamma}$ leads to another copy of the same superconformal
algebra
for the other chirality.

The above treatment gives a physical explanation for the relation between
the
superconformal algebras and superalgebras found algebraically in
\cite{Frad2,Frad3,Bowcock}.
The superconformal algebra is the boundary symmetry of an AdS supergravity
theory in the Chern-Simons formulation with the superalgebra as a gauge
group.

\section{Deriving the super-Liouville action from AdS supergravity}
\setcounter{equation}{0}

\subsection{Super-WZW model}

Because $3$-dimensional supergravity possesses the superconformal
algebra as asymptotic symmetry algebra, the dynamical theory
describing its boundary degrees of freedom at infinity is
expected to be superconformal.  We show in this section
that this is indeed the case and that the dynamical theory in question
is (extended) super-Liouville theory.  The procedure follows the
approach of \cite{CHvD} for gravity and we shall thus describe
here the main lines, emphasizing only the fine points (for related
information
concerning the relation between $AdS_3$-gravity and Liouville theory,
see \cite{NN,Ba}).

The boundary conditions (\ref{asymp}) and (\ref{asymp2})
on the superconnections $\Gamma$ and $\widetilde{\Gamma}$
can be separated into two subsets:
\begin{enumerate}
\item First, the $v$-components $\Gamma_v$ ($v \equiv x^-$)
and the $u$-components $\widetilde{\Gamma}_u$
of the superconnections along all
the generators of the superalgebra are zero,
\begin{equation}
\Gamma_v =0, \; \widetilde{\Gamma}_u =0
\label{bc1}
\end{equation}
\item Second, some of the $u$-components $\Gamma_u$ and the
$v$-components $\widetilde{\Gamma}_v$ are constrained.  More precisely,
in terms of the redefined superconnections $\Delta_i$, one has
\begin{equation}
\Delta_u^- = 1, \; \; \Delta_u^3 = 0, \; \; \Delta_{u, \, -\alpha} = 0;
\; \; \widetilde{\Delta}_v^+ = 1, \; \; \widetilde{\Delta}_v^3 = 0,
\; \; \widetilde{\Delta}_{v, \, +\alpha} = 0.
\end{equation}
It turns out that the analysis proceeds in much the same way if
the currents $\Delta_u^-$ and $\widetilde{\Delta}_v^+$ are
fixed to arbitrary non-vanishing values.  So, we shall relax
$\Delta_u^-$ and $\widetilde{\Delta}_v^+$ from their anti-de Sitter
values and consider the more general conditions
\begin{equation}
\Delta_u^- = \mu, \; \; \Delta_u^3 = 0, \; \; \Delta_{u, \, -\alpha} = 0;
\; \; \widetilde{\Delta}_v^+ = \nu, \; \; \widetilde{\Delta}_v^3 = 0,
\; \; \widetilde{\Delta}_{v, \, +\alpha} = 0.
\label{bc2}
\end{equation}
\end{enumerate}
We shall first take care of the conditions (\ref{bc1}).

It has been shown in \cite{MS,MSetAl} that pure Chern-Simons
theory on a manifold with a boundary is equivalent to
a chiral WZW theory living on that boundary under
conditions analogous to (\ref{bc1}) (the analysis of
\cite{MS,MSetAl} is straightforwardly adapted to
cover the boundary conditions (\ref{bc1}), see
\cite{CHvD}).  In the present case,
one gets two $G$-super-WZW theories on the cylinder $(t,\theta)$
at infinity, of opposite chiralities since the boundary conditions
on  $\Gamma$ and $\widetilde{\Gamma}$ are themselves
of opposite parities.  These two copies can be combined to
yield the vector $G$-super-WZW theory coupled to zero modes
whose explicit form depends on the topology of the manifold\footnote{As a
rule,
we shall not write explicitly  the zero mode terms in this section,
so the action given below describes only part of the dynamics.  This is
sufficient for exhibiting the dynamical origin of the superconformal
algebra at infinity.
Note that the zero modes
contain in particular holonomy terms, which are not zero even for the
anti-de Sitter ground state since we are using the spinor representation
of $sl_2$ \cite{BBCHO}. A discussion on how to include the
holonomies is given in the appendix.}.

Using the Gauss decomposition for a general supergroup
element,
\begin{equation}
g=E^{+}E^{0}E^{-}
\end{equation}
where
\begin{eqnarray}
E^{+}  &  =& \exp(x\sigma^{+}+\psi_{+\alpha}R^{+\alpha})=\exp(x\sigma^{+}%
)\exp(\psi_{+\alpha}R^{+\alpha})\\
E^{0}  &  =& \exp(\varphi\sigma^{3}+C^{a}T^{a})=\exp(\varphi\sigma^{3}%
)\exp(C^{a}T^{a})\\
E^{-}  &  =& \exp(y\sigma^{-}+\psi_{-\alpha}R^{-\alpha})=\exp(y\sigma^{-}%
)\exp(\psi_{-\alpha}R^{-\alpha})
\end{eqnarray}
as well as the Polyakov-Wiegmann identity \cite{PW}
for supermatrices, one gets the
WZW action
\begin{eqnarray}
S^{WSW}\left(  g\right)   &=&
\frac{k}{2\pi} \int dx^{+}dx^{-}\{\partial_{+}
\varphi\partial_{-}\varphi \nonumber \\
&+& e^{-2\varphi}\left(  \partial_{+}y-i\psi_{-\alpha
}\frac{\eta^{\alpha\beta}}{2}\partial_{+}\psi_{-\beta}\right)  \left(
\partial_{-}x-i\psi_{+\alpha}\frac{\eta^{\alpha\beta}}{2}\partial_{-}
\psi_{+\beta}\right)  \nonumber\\
&-& ie^{-\varphi}\partial_{-}
\psi_{+\alpha} u^{\alpha\beta}
\partial_{+}\psi_{-\beta}\}-\frac{2D}{d\left(  d-1\right)  }WZW[u]
\label{WZW}
\end{eqnarray}
where $u = \exp[C^a \lambda^a]$ and where
$ WZW[u]$ is the standard WZW action for $u$:
\begin{equation}
WZW[u] = \frac{k}{4 \pi} \int dx^{+}dx^{-} Tr(\partial_{-} u \,
u^{-1}
\partial_{+} u \, u^{-1}) + \Gamma[u]
\end{equation}
with $\Gamma[u]$  the WZW term.

More precisely, one gets (\ref{WZW}) in Hamiltonian form, i.e.,
\begin{eqnarray}
&{}& S_H^{WZW}[\varphi,x,y,u, \psi_{+ \alpha},
\psi_{- \alpha}, \Pi_\varphi, \Pi_x,
\Pi_y, \Pi_u, \Pi^{- \alpha}, \Pi^{+ \alpha}] \nonumber \\
&{}& \; \; \; = \int dx^{+}dx^{-} [\Pi_\varphi
\dot{\varphi} + \partial_- x \Pi_x
+ \partial_+ y \Pi_y +  \partial_+\psi_{-\beta} \Pi^{-\beta}
+ \partial_-\psi_{+\beta} \Pi^{+\beta} \nonumber \\
&{}& \; \; \; - \frac{1}{2}\left(\frac{\pi}{k} \Pi_\varphi^2
+ \frac{k}{\pi} \varphi^{'2}\right) -
\frac{2 \pi}{k} \Pi_x \Pi_y e^{2 \varphi} - \frac{ik}{2 \pi} e^{-\varphi}
\Phi_{-\alpha}u^{\alpha \beta} \Phi_{+ \beta}] \nonumber \\
&{}& \; \; \; -\frac{2D}{d\left(  d-1\right)  }WZW^H[u].
\label{HamWZW}
\end{eqnarray}
were we introduced the notations:
\begin{eqnarray} 
\Phi_{+\alpha} & =&  e^{\varphi}\left(
\frac{2\pi}{ik}\Pi_{-\beta}-\frac{\pi}{k}\Pi_y\psi_{-\beta} \right)
(u^{-1})^{\beta}_{\alpha}  \; \;
\hbox{ ($= \partial_{-}\psi_{+\alpha}$ on-shell)}\\
\Phi_{-\alpha} &=&  e^{\varphi}\left(
-\frac{2\pi}{ik}\Pi_{+\beta}+\frac{\pi}{k}\Pi_x\psi_{+\beta} \right)
u^{\beta}_{\alpha}  \; \;
\hbox{ ($= \partial_{+}\psi_{-\alpha}$ on-shell)} 
\end{eqnarray}

Indeed, the chiral actions are in first-order form, and one goes
from the sum of the chiral actions to (\ref{HamWZW}) by making a change
of phase space variables \cite{CHvD,Lys}.

If one eliminates the conjugate momenta from (\ref{HamWZW})
by using their own equations of motion, one gets (\ref{WZW}).
The relationship between the conjugate momenta and the temporal
derivatives
of
the fields is  explicitly given by  (recall that $x^\pm = t \pm \theta$,
so that $2 \partial_\pm = \partial_t \pm \partial_\theta$ and
$dx^+ dx^- = 2 dt d \theta$)
\begin{eqnarray}
\Pi_\varphi &=& \frac{k}{2\pi} \dot{\varphi} \\
\Pi_{x} &=& \frac{k}{2 \pi}e^{-2\varphi}\left(  \partial_{+}y
-i\psi_{-\alpha}\frac{\eta^{\alpha\beta}}{2}\partial_{+}\psi_{-\beta}
\right)  \label{moment1}\\
\Pi_{y} &=& \frac{k}{2 \pi}e^{-2\varphi}\left(  \partial_{-}
x-i\psi_{+\alpha}\frac{\eta^{\alpha\beta}}{2}\partial_{-}\psi_{+\beta}
\right) \label{moment2}
\end{eqnarray}
and
\begin{eqnarray}
\Pi^{+\beta} &=& \frac{\partial^L L}{\partial \dot{\psi}_{+ \beta}}
\nonumber \\
&=&  \frac{i k}{2 \pi}
[ \frac{1}{2} e^{-2\varphi}\left(  \partial_{+}y
-i\psi_{-\gamma}\frac{\eta^{\gamma\delta}}{2}\partial_{+}\psi_{-\delta}
\right)  \psi_{+\alpha} \eta^{\alpha \beta} \nonumber \\
& & \; \; \; \; \;  -e^{-\varphi}
u^{\beta \alpha}
\partial_{+}\psi_{-\alpha}],
\end{eqnarray}
\begin{eqnarray}
\Pi^{-\beta} &=& \frac{\partial^L L}{\partial \dot{\psi}_{- \beta}}
\nonumber \\
&=&  \frac{i k}{2 \pi}
[ \frac{1}{2} e^{-2\varphi}\left(  \partial_{-}x
-i\psi_{+\gamma}\frac{\eta^{\gamma\delta}}{2}\partial_{-}\psi_{+\delta}
\right)  \psi_{-\alpha} \eta^{\alpha \beta} \nonumber \\
& & \; \; \; \; \;  +e^{-\varphi}
\left(u^{-1}\right)^{\beta \alpha}
\partial_{-}\psi_{+\alpha}]
\end{eqnarray}

\subsection{Kac-Moody currents}

As it
is well known , the super WZW action has two sets (``left'' and
``right'')
of conserved currents, each set forming a super Kac-Moody
algebra.  With appropriate normalizations,
the currents for the two chiralities can be taken to be
$\frac{k}{2\pi}\partial_{+}gg^{-1}$ and $-\frac{k}{2\pi}
g^{-1}\partial_{-}g$ and we define the
current components by
\begin{equation}
\frac{k}{2\pi} \partial_{+}gg^{-1}
\equiv a^{I} G^{I}
\equiv \left(  J^{3}\frac{\sigma^{3}}{2}
+J^{+}\sigma^{+}+J^{-}\sigma^{-}\right)  +\left( \frac{d-1}{2C_{\rho}}
B^{a}T^{a}\right)
+\left(  F_{+\alpha}R^{+\alpha}+F_{-\alpha}R^{-\alpha}\right)
\end{equation}
\begin{equation}
-\frac{k}{2\pi} g^{-1}\partial_{-}g
\equiv \widetilde{a}^I G^{I}
\equiv \left(  \widetilde{J}^{3}\frac{\sigma^{3}}{2}
+\widetilde{J}^{+}\sigma^{+}+\widetilde{J}^{-}\sigma^{-}\right)  +\left(
\frac{d-1}{2C_{\rho}}
\widetilde{B}^{a}T^{a}\right)
+\left(
\widetilde{F}_{+\alpha}R^{+\alpha}+\widetilde{F}_{-\alpha}R^{-\alpha}\right)
\end{equation}
where $a^{I},\widetilde{a}^I$ are generically the components of the
currents
and $G^{I}$ the representation matrices of the superalgebra.
The super K-M Poisson algebra (with $\{q,p\} = 1$)
between any two current components is then:
\begin{equation}
\{a^{I}(\theta),a^{J}(\theta')\}=g^{IJK}a^K(\theta)\delta(\theta-\theta')
+\frac{k}{2\pi}\,h^{IJ}\, \delta'(\theta-\theta') \end{equation} 
\begin{equation}
\{\widetilde{a}^I(\theta),\widetilde{a}^J(\theta')\}
=g^{IJK}\widetilde{a}^K(\theta)\delta(\theta-\theta')
-\frac{k}{2\pi}\,h^{IJ}\, \delta'(\theta-\theta') \end{equation}
where
\begin{equation} \left[ G^{I},G^{J} \right]=f^{IJK}G^{K},\quad STr \left(
G^{I}G^{J} \right) = g^{IJ} \end{equation} and the commutator of two
superalgebra generators is understood as an anti-commutator for two odd
generators.  We have also defined $h^{IJ}$ to be the inverse
matrix to $g^{IJ}$, $h^{IJ} g^{JM}= \delta^{IM}$ and
$g^{IJK} = f^{LMN} h^{IL} h^{JM} g^{NK}$.

In terms of the fields, the currents for the two chiralities read
explicitly:

\begin{equation}
J^3 = \frac{k}{\pi}\partial_+ \varphi +2\Pi_x x -
\Pi^{+\alpha}\psi_{+\alpha} 
\end{equation}
\begin{equation}
J^- = \Pi_x
\end{equation}
\begin{eqnarray}
& &{} J^+  = - \Pi_{y}\left[ -e^{2\varphi} +\frac{i}{2}
e^{\varphi}\left( \psi_{+\alpha}u^{\alpha\beta}\psi_{-\beta}
\right) \right] +e^{\varphi}\left(
\psi_{+\alpha}u^{\alpha\beta}\Pi_{-\beta} \right)
\nonumber\\& & \; \; \; \; \; \; \; +\frac{k}{2\pi}\left[
\partial_{\theta}x
+\frac{i}{2}\left( \psi_{+\alpha}\eta^{\alpha\beta}
\partial_{\theta}\psi_{+\beta} \right)
-2x\partial_{+}\varphi
-\frac{i}{2}\left( \partial_{+}u u^{-1} \right)^a
\left( \psi_{+\alpha}(\lambda^a)^{\alpha\beta}\psi_{+\beta} \right)
\right]
\nonumber \\
& & \; \; \; \; \; \; \; -\Pi_{x}\left[ x^2+\frac{1}{4!}
\frac{d-1}{2C_{\rho}}
\left( \psi_{+\alpha}(\lambda^a)^{\alpha\beta}\psi_{+\beta} \right)
\left( \psi_{+\gamma}(\lambda^a)^{\gamma\delta}\psi_{+\delta}
\right)
\right] \nonumber\\
& & \; \; \; \; \; \; \; -\left[
\frac{1}{2}\Pi_{x}\psi_{+\beta}+i\Pi^{+\alpha}\eta_{\alpha\beta}
\right]
\left[ ix\eta^{\beta\gamma}\psi_{+\gamma}-\frac{1}{3!}
\frac{d-1}{2C_{\rho}}
\left( \lambda^a \right) ^{\beta\gamma} \psi_{+\gamma}
\left( \psi_{+\alpha}(\lambda^a)^{\alpha\beta}\psi_{+\beta} \right)
\right]
\end{eqnarray} 
\begin{equation}
B^a = \frac{k}{2\pi} \frac{2C_{\rho}}{d-1} \left( \partial_+ uu^{-1}
\right) ^a - \Pi_{+\alpha}(\lambda^a)^{\alpha\beta}\psi_{+\beta}
\end{equation}
\begin{equation}
F_{-\alpha} = i\eta_{\alpha\beta}\Pi^{+\beta}-\frac{1}{2}\Pi_x\psi_{+\alpha}
\end{equation}
\begin{eqnarray}
& & F_{+\alpha}  =  -e^{\varphi}
\left[
\frac{1}{2}\psi_{-\gamma}\eta^{\gamma\beta}\Pi_{y}+i\Pi^{-\beta}
\right](u^{-1})_{\beta\alpha}{}
\nonumber\\
& & \; \; \; \; \; \; \; +\frac{k}{2\pi}\left[
\partial_{\theta}\psi_{+\alpha}
-\partial_{+}\varphi\psi_{+\alpha}
-\left( \partial_{+}u u^{-1} \right)^a
\left( \psi_{+\beta}\left( \lambda^a \right)
^{\beta}_{\alpha}
\right) \right] {}
\nonumber\\
& & \; \; \; \; \; \; \; +\Pi_{x}\left[ -x\psi_{+\alpha}-\frac{i}{3!}
\frac{d-1}{2C_{\rho}}
\left( \psi_{+\gamma}(\lambda^a)^{\gamma\beta}\psi_{+\beta} \right)
\left( \psi_{+\delta}(\lambda^a)^{\delta}_{\alpha} \right)
\right] {}
\nonumber\\
& & \; \; \; \; \; \; \; +\left[
\frac{1}{2}\Pi_{x}\psi_{+\delta}\eta^{\delta\beta}+i\Pi^{+\beta}
\right]
\left[ x-\frac{i}{4}\psi_{+\alpha}\psi_{+\beta}
+\frac{i}{2!}\frac{d-1}{2C_{\rho}}
\left( (\lambda^a)^{\beta\gamma}\psi_{+\gamma} \right)
\left( \psi_{+\delta}(\lambda^a)^{\delta}_{\alpha} \right)
\right]
\end{eqnarray}
and for the other chirality:
\begin{equation}
\widetilde{J}^3  = - \frac{k}{\pi}\partial_{-}\varphi 
         -2\Pi_y y +\Pi^{-\alpha}\psi_{-\alpha}
\end{equation}
\begin{equation}
\widetilde{J}^{+}  = - \Pi_{y}
\end{equation}
\begin{eqnarray}
& &{} \widetilde{J}^-  =   \Pi_{x}\left[ -e^{2\varphi} +\frac{i}{2}
e^{\varphi}\left( \psi_{+\alpha}u^{\alpha\beta}\psi_{-\beta}
\right) \right] -e^{\varphi}\left(
\Pi_{+\alpha}u^{\alpha\beta}\psi_{-\beta} \right)
\nonumber\\& & \; \; \; \; \; \; \; +\frac{k}{2\pi}\left[
\partial_{\theta}y
+\frac{i}{2}\left( \psi_{-\alpha}\eta^{\alpha\beta}
\partial_{\theta}\psi_{-\beta} \right)
+2y\partial_{-}\varphi
-\frac{i}{2}\left( u^{-1}\partial_{-}u \right)^a
\left( \psi_{-\alpha}(\lambda^a)^{\alpha\beta}\psi_{-\beta} \right)
\right]
\nonumber \\
& & \; \; \; \; \; \; \; +\Pi_{y}\left[ y^2+\frac{1}{4!}
\frac{d-1}{2C_{\rho}}
\left( \psi_{-\alpha}(\lambda^a)^{\alpha\beta}\psi_{-\beta} \right)
\left( \psi_{-\gamma}(\lambda^a)^{\gamma\delta}\psi_{-\delta}
\right)
\right] \nonumber\\
& & \; \; \; \; \; \; \; +\left[
\frac{1}{2}\Pi_{y}\psi_{-\beta}+i\Pi^{-\alpha}\eta_{\alpha\beta}
\right]
\left[ iy\eta^{\beta\gamma}\psi_{-\gamma}-\frac{1}{3!}
\frac{d-1}{2C_{\rho}}
\left( \lambda^a \right) ^{\beta\gamma} \psi_{-\gamma}
\left( \psi_{-\alpha}(\lambda^a)^{\alpha\beta}\psi_{-\beta} \right)
\right]
\end{eqnarray}
\begin{equation}
\widetilde{B}^a = -\left[ \frac{k}{2\pi} \frac{2C_{\rho}}{d-1}\left(
u^{-1}\partial_{-}u \right)^a
+\left( \Pi_{-\alpha}(\lambda^a)^{\alpha\beta}\psi_{-\beta} \right)
\right]
\end{equation}
\begin{equation}
\widetilde{F}_{+\alpha} = i \eta_{\alpha\beta} \Pi^{-\beta} -\frac{1}{2}
\Pi_{y}\psi_{-\alpha} \end{equation}
\begin{eqnarray}
& & \widetilde{F}_{-\alpha}  =  -e^{\varphi}
\left[
\frac{1}{2}\psi_{+\gamma}\eta^{\gamma\beta}\Pi_{x}+i\Pi^{+\beta}
\right]u_{\beta\alpha}{}
\nonumber\\
& & \; \; \; \; \; \; \; +\frac{k}{2\pi}\left[
\partial_{\theta}\psi_{-\alpha}
+\partial_{-}\varphi\psi_{-\alpha}
+\left( u^{-1}\partial_{-}u \right)^a
\left( \psi_{-\beta}\left( \lambda^a \right)
^{\beta}_{\alpha}
\right) \right] {}
\nonumber\\
& & \; \; \; \; \; \; \; -\Pi_{y}\left[ -y\psi_{-\alpha}-\frac{i}{3!}
\frac{d-1}{2C_{\rho}}
\left( \psi_{-\gamma}(\lambda^a)^{\gamma\beta}\psi_{-\beta} \right)
\left( \psi_{-\delta}(\lambda^a)^{\delta}_{\alpha} \right)
\right] {}
\nonumber\\
& & \; \; \; \; \; \; \; -\left[
\frac{1}{2}\Pi_{y}\psi_{-\delta}\eta^{\delta\beta}+i\Pi^{-\beta}
\right]
\left[ y-\frac{i}{4}\psi_{-\alpha}\psi_{-\beta}
+\frac{i}{2!}\frac{d-1}{2C_{\rho}}
\left( (\lambda^a)^{\beta\gamma}\psi_{-\gamma} \right)
\left( \psi_{-\delta}(\lambda^a)^{\delta}_{\alpha} \right)
\right]
\end{eqnarray}

\subsection{Gauged theory}

The variational principle that follows from $2+1$
supergravity is in fact not (\ref{HamWZW}), but rather,
(\ref{HamWZW}) supplemented by the constraints (\ref{bc2})
since the only competing histories occuring in the variational
principle of supergravity are required to fulfill (\ref{bc2}).
In terms of the currents, the constraints are simply
\begin{eqnarray}
J^- = \frac{k}{2 \pi} \mu &,& \widetilde{J}^+ = \frac{k}{2 \pi} \nu
\label{1stclass}
\\
J^3 = 0  &,& \widetilde{J}^3 = 0 \label{2ndclass} \\
F_{- \alpha} = 0 &,& \widetilde{F}_{+ \alpha} = 0
\label{mixedclass}
\end{eqnarray}
(modulo  holonomy terms discussed in the appendix).
These are constraints on the canonical variables (with $\partial_\pm
\varphi$ expressed
in terms of $\Pi_\varphi$),
which we collectively denote by  $G_A$.
Thus, the action for the theory is
\begin{eqnarray}
&{}& S[\varphi,x,y,u,\psi_{+\alpha}, \psi_{-\alpha}, \Pi_\varphi,
\Pi_x,\Pi_y, \Pi_u, \Pi^{-\alpha},\Pi^{+ \alpha}, \xi^A] \nonumber \\
&{}& \; \; \; \; \; \; \; \; \; \; \; \;
= S^{WZW}_H - \int dx^+ dx^- \xi^A G_A
\label{action0}
\end{eqnarray}
where $\xi^A$ are Lagrange multipliers implementing the constraints.

It follows from the algebra of the currents that
the constraints $G_A = 0$ are second class; one may split them into
two subsets, $G_A = (\phi_i, \chi_i)$, in such a way that the
$\phi_i$'s are first class among themselves and generate therefore
a gauge symmetry,
while the $\chi_i$'s can be viewed as gauge
conditions for the symmetry generated by the $\phi_i$'s
(for information on constrained systems,
see \cite{HT}).
For instance, one may take for $\phi_i$
the constraints $J^- - \frac{k}{2 \pi} \mu = 0$, $\widetilde{J}^+
- \frac{k}{2 \pi} \nu = 0$ and the positive frequency part of the
fermionic constraints \cite{BO}.
The action (\ref{action0}) is thus the (gauge-fixed
version of the) action
for the gauged WZW model, in which one has gauged the
subsupergroup generated by
the first class constraints.

As it is well known, this model is equivalent to super-Liouville
theory (see \cite{X,XX,XXX} for the bosonic case, and \cite{BO,Z,ZZ}
for some of the fermionic cases).
This is demonstrated in the next subsection for all extended
models.

\subsection{Super-Liouville action}

The constraints $G_A = 0$ enable one to eliminate $\Pi_x,\Pi_y,
\Pi^{-\alpha},\Pi^{+ \alpha}, x,y$ from the action.
[As demonstrated by Lagrange himself, solving the constraints
inside the action or taking them into account with the help of
(Lagrange) multipliers
are two equivalent procedures].
When doing so, one gets a reduced action $S^R_H$ which depends on the
remaining variables, i.e., $\varphi, u,\psi_{+\alpha}, \psi_{-\alpha},
\Pi_\varphi,
\Pi_u$ and which is precisely the super-Liouville action.

Indeed, the constraints imply
\begin{equation}
\Pi_x =  \frac{k}{2 \pi} \mu, \; \; \Pi_y = - \frac{k}{2 \pi} \nu,
\end{equation}
and
\begin{equation}
\Pi^{+\beta} = -\frac{ik\mu}{4 \pi} \eta^{\beta \alpha} \psi_{+\alpha},
\; \; \Pi^{-\beta} =
\frac{ik\nu}{4 \pi} \eta^{\beta \alpha} \psi_{-\alpha}
\end{equation}
When substituting this inside (\ref{action0}), one
gets, dropping a total derivative
and integrating over the remaining momenta $\pi_\varphi$
and $\Pi_u$,
\begin{eqnarray}
S_{SL}&=&\frac{k}{2\pi}\int  dx^{+}dx^{-}\left\{
\begin{array}
[c]{c}
\partial_{+}\varphi\partial_{-}\varphi +\mu\nu \left( 
e^{2\varphi}-ie^{\varphi}\psi_{+\alpha} u^{\alpha\beta}\psi_{-\beta} \right)\\
+\frac{i\mu}{2}\psi_{+\alpha}\eta^{\alpha\beta}\partial_{-}\psi_{+\beta}
-\frac{i\nu}{2}\psi_{-\alpha}\eta^{\alpha\beta}\partial_{+}\psi_{-\beta}
\end{array}
\right\} 
\nonumber \\
&-&\frac{2D}{d\left(  d-1\right)  }WZW(u)
\end{eqnarray}
which is just the super-Liouville action
\cite{KPR,ZH}.

The values of $\mu$, $\nu$ corresponding to the anti-de Sitter case
are $\mu = 1$ and $\nu = 1$, leading to the action
\begin{eqnarray}
S_{SL}&=&\frac{k}{2\pi}\int  dx^{+}dx^{-}\left\{
\begin{array}
[c]{c}
\partial_{+}\varphi\partial_{-}\varphi + \left(
e^{2\varphi}-ie^{\varphi}\psi_{+\alpha} u^{\alpha\beta}\psi_{-\beta} \right)\\
+\frac{i}{2}\psi_{+\alpha}\eta^{\alpha\beta}\partial_{-}\psi_{+\beta}
-\frac{i}{2}\psi_{-\alpha}\eta^{\alpha\beta}\partial_{+}\psi_{-\beta}
\end{array}
\right\}
\nonumber \\
&-&\frac{2D}{d\left(  d-1\right)  }WZW(u).
\label{superLiouvilleAction}
\end{eqnarray}
The sign in front of the exponential is not the one
familiar from $2D$ gravity.   This point is discussed more
fully in the appendix.

\section{Realization of superconformal generators in terms of Liouville
fields}
\setcounter{equation}{0}

\subsection{Construction of generators}

The symmetries of the super-Liouville action can be understood
in terms of the symmetries of
the original, ungauged, WZW action.

The superWZW model is conformal, the generators $L$,$\widetilde{L}$ being
elements of the enveloping algebra of the current algebra defined by
the Sugawara construction:
\begin{equation}
\frac{1}{2\pi}L=\frac{1}{k}\left((\frac{J^3}{2})^2+J^{+}J^{-} \right)
 +\frac{1}{k}\eta^{\alpha\beta}F_{+\alpha}F_{-\beta} +\frac{d-1}{2C_\rho}
\frac{B^aB^a}{2k} \end{equation}
\begin{equation}
- \frac{1}{2\pi}\widetilde{L}=\frac{1}{k}\left((\frac{
\widetilde{J}^3}{2})^2+\widetilde{J}^{+}\widetilde{J}^{-} \right)
+\frac{1}{k}\eta^{\alpha\beta}\widetilde{F}_{+\alpha}\widetilde{F}_{-\beta}
+\frac{d-1}{2C_\rho}
\frac{\widetilde{B}^a\widetilde{B}^a}{2k} \end{equation}

The  Poisson brackets of $L$,$\widetilde{L}$ with the currents,  measuring
their dimension
are: \begin{equation}
\{L(\theta),a^I(\theta')\}=a^I(\theta)\delta'(\theta-\theta')
\end{equation}
\begin{equation}
\{\widetilde{L}(\theta),\widetilde{a}^I(\theta')\}=\widetilde{a}^I(\theta)
\delta'(\theta-\theta')
\end{equation}
The ungauged superWZW model is, however, not superconformal, in the sense
that there
are in general no superpartners to $L$,$\widetilde{L}$ that would close
with them according
to the superconformal algebra.  What is remarkable is that once we gauge
the superWZW, i.e., impose the first class constraints
\begin{equation}
  J^{-}=\frac{k}{2\pi}\mu, \,\,\,\,\,\,\,\,\,\,F^{>}_{-\alpha}=0,
\,\,\,\,\,\,\,\,\,\,
\widetilde{J}^{+}=\frac{k}{2\pi}\nu,\,\,\,\,\,\,\,\,\,\,
\widetilde{F}^{>}_{+\alpha}=0,
  \label{constraints}
\end{equation}
(where $>$ denotes the positive frequency part),
we not only maintain conformal invariance but in fact gain superconformal
invariance.  Indeed, one can find polynomials in the Kac-Moody currents
that (i) preserve the constraints; and (ii) close in the
Poisson brackets according to the superconformal algebra modulo
terms that vanish when (\ref{constraints}) hold.

In the reduced theory obtained by strongly enforcing all the constraints
(gauge constraints (\ref{constraints}) and
gauge conditions $J^3 = 0$, $\widetilde{J}^3 = 0$, $F^{<}_{+\alpha}=0$,
$\widetilde{F}^{<}_{-\alpha}=0$)
and using Dirac brackets,
the superconformal algebra is preserved since the generators are
``first-class" (=gauge-invariant) so that their
Dirac and Poisson brackets coincide.

We exemplify the procedure by constructing the superconformal generators
$Q_{\alpha}$,$\widetilde{Q}_{\alpha}$.  In the actual construction, it is
convenient to work in a
``half-gauge-fixed"
formulation in which $F^{<}_{+\alpha}=0$, $\widetilde{F}^{<}_{-\alpha}=0$ are
imposed, so that the
constraints
are
\begin{equation}
  J^{-}=\frac{k}{2\pi}\mu, \,\,\,\,\,\,\,\,\,\,F_{-\alpha}=0.
\end{equation}
\begin{equation}
  \widetilde{J}^{+}=\frac{k}{2\pi}\nu,
\,\,\,\,\,\,\,\,\,\,\widetilde{F}_{+\alpha}=0.
    \label{constraints'}
\end{equation}
One gets for $Q_{\alpha}$
\begin{equation}
\frac{k\mu}{2\pi} Q_{\alpha}\equiv J^{-}F_{+\alpha}-\frac{J^3}{2} 
F_{-\alpha}
+\frac{d-1}{2C_{\rho}}B^a \eta^{\beta\gamma} \left( \lambda^a
\right)_{\alpha\beta}F_{-\gamma}
+\frac{k}{2\pi} \partial_{\theta}F_{-\alpha}(\theta)
\label{QQ}           
\end{equation}
and for $\widetilde{Q}_{\alpha}$:
\begin{equation}
\frac{k\nu}{2\pi} \widetilde{Q}_{\alpha}\equiv
\widetilde{J}^{+}\widetilde{F}_{-\alpha}+\frac{\widetilde{J}^3}{2}
\widetilde{F}_{+\alpha} +\frac{d-1}{2C_{\rho}}\widetilde{B}^a
\eta^{\beta\gamma} 
\left(\lambda^a \right)_{\alpha\beta}\widetilde{F}_{+\gamma
}-\frac{k}{2\pi} \partial_{\theta}\widetilde{F}_{+\alpha}(\theta)
\label{QQ2}
\end{equation}    
The Poisson brackets of  $Q_{\alpha}$,$\widetilde{Q}_{\alpha}$ with the
constraints are:
\begin{equation}
\{Q_{\alpha}(\theta),J^{-}(\theta') -\frac{k}{2 \pi} \mu\}=0, \; \; \;
\{\widetilde{Q}_{\alpha}(\theta),\widetilde{J}^{+}(\theta')
- \frac{k}{2 \pi} \nu \}=0
\end{equation}
and
\begin{eqnarray}
\frac{k\mu}{2\pi}\{Q_{\alpha}(\theta),F_{-\beta}(\theta')\}
& = & \frac{k}{2\pi}\eta_{\alpha \beta}
\delta(\theta-\theta') \partial_ \theta
 J^-(\theta) + \frac{1}{2} F_{-\beta}
 (\theta)F_{-\alpha}(\theta)\delta(\theta
 - \theta') +{}
 \nonumber\\
& & {}-\frac{d-1}{2C_\rho}\left( \lambda^a \right)_{\alpha}^{\epsilon}
 \left( \lambda^a \right)_{\beta}^{\tau}
  F_{-\tau}(\theta)F_{-\epsilon}(\theta)\delta(\theta-\theta'),
\end{eqnarray}                 
\begin{eqnarray}
\frac{k\nu}{2\pi}\{\widetilde{Q}_{\alpha}(\theta),
\widetilde{F}_{+\beta}(\theta')\}
& = & \frac{k}{2\pi}\eta_{\alpha \beta}
\delta(\theta-\theta') \partial_ \theta
 \widetilde{J}^+(\theta) - \frac{1}{2} \widetilde{F}_{+\beta}
 (\theta)\widetilde{F}_{+\alpha}(\theta)\delta(\theta
 - \theta') +{}
 \nonumber\\
& & {}-\frac{d-1}{2C_\rho}\left( \lambda^a \right)_{\alpha}^{\epsilon}
 \left( \lambda^a \right)_{\beta}^{\tau}
\widetilde{F}_{+\tau}(\theta)
\widetilde{F}_{+\epsilon}(\theta)\delta(\theta-\theta'),
\end{eqnarray}          
expressions that vanish on the constraint surface.
Note that the
dimension of $Q_{\alpha}$,$\widetilde{Q}_{\alpha}$ are the difference
between the dimension
as measured by the Sugawara tensor and the the $sl(2,R)$ spin.

The superconformal generators, as defined above, generate the
superconformal algebra . Their Poisson brackets close on modified
generators, all having (weakly)
vanishing Poisson brackets with the constraints
(the bracket of two first class functions is also first class).
In particular the modified Virasoro generators
$\widehat{L}$,$\widehat{\widetilde{L}}$ which appear are
\begin{equation}
  \frac{k\mu}{2\pi}\widehat{L} \equiv
-J^-(\theta)(L(\theta)-\partial_{\theta} J^3/2 ), \;
  \frac{k\nu}{2\pi}\widehat{\widetilde{L}} \equiv
-\widetilde{J}^+(\theta)(\widetilde{L}(\theta)+\partial_{\theta}
\widetilde{J}^3/2 )
  \label{LL}
\end{equation}     
Their Poisson brackets with the constraints are:
\begin{eqnarray}
\{\widehat{L}(\theta),J^-(\theta')
-\frac{k}{2 \pi} \mu\} &=& -J^-(\theta)
\partial_{\theta}J^-(\theta)\delta(\theta-\theta') \\
\{\widehat{\widetilde{L}}(\theta),\widetilde{J}^{+}(\theta')
- \frac{k}{2 \pi} \nu\}
&=& \widetilde{J}^{+}(\theta)\left(
\partial_{\theta} \widetilde{J}^{+}(\theta) \right)
\delta(\theta-\theta') \\
\{\widehat{L}(\theta),F_{-\alpha}(\theta')\} &=&-\frac{1}{2}J^-(\theta)
\left[ F_{-\alpha}(\theta)\delta'(\theta-\theta')
-\partial_{\theta}F_{-\alpha}(\theta)\delta(\theta-\theta')\right]  
\\
\{ \widehat{\widetilde{L}}(\theta),\widetilde{F}_{+\alpha}(\theta') \}
&=&-\frac{1}{2}\widetilde{J}_{+}(\theta)\left[
\widetilde{F}_{+\alpha}(\theta)\delta'(\theta-\theta')-\partial_{\theta}
\widetilde{F}_{+\alpha}(\theta) \delta(\theta-\theta') \right]
\end{eqnarray}        
again vanishing  on the manifold of the constraints.
Similarly the modified Kac-Moody currents
$-\frac{2\pi}{k\mu}J_-(\theta)B^a(\theta)$ and 
$-\frac{2\pi}{k\nu}\widetilde{J}_+ (\theta)\widetilde{B}^a(\theta)$
appear.

The algebraic structure obtained is indeed the extended
superconformal algebra provided we ignore terms which vanish with the
constraints and we consider as equivalent, generators which reduce to
the same expression on the constraint
surface.  This is of course the standard procedure for
constrained systems and it lies at the heart of the
Hamiltonian formulation of BRST theory \cite{HT}.

When all the constraints (first and second class)
are imposed, the generator $Q_{\alpha}$ reduces to
$F_{+\alpha}$ and $\widetilde{Q}_{\alpha}$ reduces to
$\widetilde{F}_{-\alpha}$, while the generators $\widehat{L}$ and
$\widetilde{\widehat{L}}$ reduce to 
$-\mu J^{+} -\frac{\pi} 
{k_B} B^aB^a $ and
$+\nu \widetilde{J^-} + \frac{\pi}
{k_B}\widetilde{B}^a \widetilde{B}^a
$ respectively.
Therefore, one can also view
the superconformal generators as the ``first-class extensions" \cite{HT}
of the Kac-Moody currents that are not constrained by the reduction.

\subsection{Explicit expressions}

One can directly obtain the superconformal generators in terms of the
super-Liouville fields
by plugging the constraints into the above expressions (\ref{QQ}),
(\ref{LL}).

On the constraint surface, the currents reduce to
\begin{eqnarray}
J^3 & =&  \frac{k}{\pi} \left[ \partial_{+}\varphi +\mu x \right] \\
J^{-} & = & \frac{k}{2\pi} \mu \\
J^{+} & = &  \frac{k}{2\pi} [ -\nu e^{2\varphi}
+i\nu e^{\varphi}\left( \psi_{+\alpha} u^{\alpha\beta} \psi_{-\beta}
\right) \nonumber\\ & +& \partial_{\theta} x+\frac{i}{2}\psi_{+\alpha}
\eta^{\alpha\beta} \partial_{\theta} \psi_{+\beta}-2x\partial_{+} \varphi
-\mu x^2 \nonumber\\ & -& \frac{i}{2} \left( \partial_{+}u u^{-1} \right)
^a
  \left( \psi_{+\alpha}(\lambda^a)^{\alpha\beta}\psi_{+\beta} \right)
\nonumber\\ & +& \frac{\mu}{8}\frac{d-1}{2C_{\rho}} \left(
\psi_{+\alpha}(\lambda^a)^{\alpha\beta}\psi_{+\beta} \right) \left(
\psi_{+\gamma}(\lambda^a)^{\gamma\delta}\psi_{+\delta}\right) ] \\ 
B^a & = & \frac{k}{2\pi} \left( \frac{2C_{\rho}}{d-1}
        \left( \partial_{+}u u^{-1}\right) ^a
     +\frac{i\mu}{2}
       \left( \psi_{+\alpha}(\lambda^a)^{\alpha\beta}\psi_{+\beta} \right)
\right) \\ 
F_{-\alpha} &=& 0 \\ 
F_{+\alpha} & = & \frac{k}{2\pi} [
\nu e^{\varphi}\left( u \right) ^{\beta}_{\alpha}\psi_{-\beta}
+\partial_{\theta} \psi_{+\alpha} \nonumber\\ & -&
\partial_{+}\varphi\psi_{+\alpha} -\left( \partial_+ u u^{-1} \right) ^a
\left( \psi_{+\beta}\left( \lambda^a \right) ^{\beta}_{\alpha} \right)
\nonumber\\ & +& \frac{i\mu}{3} \frac{d-1}{2C_{\rho}} \left(
\psi_{+\gamma}(\lambda^a) ^{\gamma\beta}\psi_{+\beta} \right) \left(
\psi_{+\delta}(\lambda^a)^{\delta}_ {\alpha} \right) ] \\
\widetilde{J}^3 & =& - \frac{k}{\pi} \left[ \partial_{-}\varphi -\nu y
\right] \\
\widetilde{J}^{+} & = & \frac{k}{2\pi} \nu \\
\widetilde{J}^{-} & = &  \frac{k}{2\pi} [ -\mu e^{2\varphi}
+i\mu e^{\varphi}\left( \psi_{+\alpha}u^{\alpha\beta}\psi_{-\beta}
\right) \nonumber\\ & +& \partial_{\theta} y+\frac{i}{2}\psi_{-\alpha}
\eta^{\alpha\beta} \partial_{\theta} \psi_{-\beta}+2y\partial_{-} \varphi
-\nu y^2 \nonumber\\ & -& \frac{i}{2} \left( u^{-1} \partial_{-}u \right)^a
  \left( \psi_{-\alpha}(\lambda^a)^{\alpha\beta}\psi_{-\beta} \right)
\nonumber\\ & +& \frac{\nu}{8}\frac{d-1}{2C_{\rho}} \left(
\psi_{-\alpha}(\lambda^a)^{\alpha\beta}\psi_{-\beta} \right) \left(
\psi_{-\gamma}(\lambda^a)^{\gamma\delta}\psi_{-\delta}\right) ] \\ 
\widetilde{B}^a & = & -\frac{k}{2\pi} \left( \frac{2C_{\rho}}{d-1}
        \left( u^{-1}\partial_{-}u \right) ^a
     +\frac{i\nu}{2}
       \left( \psi_{-\alpha}(\lambda^a)^{\alpha\beta}\psi_{-\beta} \right)
\right) \\ 
\widetilde{F}_{+\alpha} &=& 0 \\ 
\widetilde{F}_{-\alpha} & = & \frac{k}{2\pi} [
\mu e^{\varphi} \psi_{+\beta} u^{\beta}_{\alpha}
+\partial_{\theta} \psi_{-\alpha} \nonumber\\ & +&
\partial_{-}\varphi\psi_{-\alpha} +\left( u^{-1} \partial_{-}u \right) ^a
\left( \psi_{-\beta}\left( \lambda^a \right) ^{\beta}_{\alpha} \right)
\nonumber\\ & +& \frac{i\nu}{3} \frac{d-1}{2C_{\rho}} \left(
\psi_{-\gamma}(\lambda^a) ^{\gamma\beta}\psi_{-\beta} \right) \left(
\psi_{-\delta}(\lambda^a)^{\delta}_ {\alpha} \right) ] \end{eqnarray}

If one inserts these expressions into (\ref{QQ}) and (\ref{LL}), one gets
the following expressions for the superconformal generators:
\begin{eqnarray}
\widehat{L} &=& \frac{k}{2\pi}\left[
               \partial_{+}^2\varphi -\left( \partial_{+}\varphi\right) ^2 
-\frac{i\mu}{2}\psi_{+\alpha}\eta^{\alpha\beta}
                 \partial_{+}\psi_{+\beta}
              -\frac{C_{\rho}}{d-1}\left( \partial_{+}u u^{-1} \right)^a
               \left( \partial_{+}u u^{-1} \right)^a \right] \label{LLL}\\   
-B^a &=& -\frac{k}{2\pi} \left[ \frac{2C_{\rho}}{d-1}
\left( \partial_{+}u u^{-1} \right)^a
+\frac{i\mu}{2}
\left( \psi_{+\alpha}(\lambda^a)^{\alpha\beta}\psi_{+\beta} \right)
\right] \label{BBB}\\   
Q_{\alpha} & = & \frac{k}{2\pi} [ \partial_{+}\psi_{+\alpha}
-\partial_{+}\varphi\psi_{+\alpha} -\left( \partial_{+}u u^{-1} \right) ^a
\left( \psi_{+\beta}\left( \lambda^a \right) ^{\beta}_{\alpha}\right)
\nonumber\\ & +& \frac{i\mu}{3}\frac{d-1}{2C_{\rho}} \left(
\psi_{+\gamma}(\lambda^a)^{\gamma\beta}\psi_{+\beta} \right) \left(
\psi_{+\delta}(\lambda^a)^{\delta}_{\alpha} \right) ] \label{QQQ}\\     
\widehat{\widetilde{L}} &=& -\frac{k}{2\pi}\left[
               \partial_{-}^2\varphi -\left( \partial_{-}\varphi\right) ^2
+\frac{i\nu}{2}\psi_{-\alpha}\eta^{\alpha\beta}
                 \partial_{-}\psi_{-\beta}
              -\frac{C_{\rho}}{d-1}\left( u^{-1}\partial_{-}u \right)^a
               \left( u^{-1}\partial_{-}u \right)^a \right] \label{LLL2}\\
-\widetilde{B}^a &=& \frac{k}{2\pi} \left[ \frac{2C_{\rho}}{d-1}
\left( u^{-1}\partial_{+}u \right)^a
+\frac{i\nu}{2}
\left( \psi_{-\alpha}(\lambda^a)^{\alpha\beta}\psi_{-\beta} \right)   
\right] \label{BBB2}\\
\widetilde{Q}_{\alpha} & = & -\frac{k}{2\pi} [ \partial_{-}\psi_{-\alpha}
-\partial_{-}\varphi\psi_{-\alpha} -\left( u^{-1}\partial_{-}u \right) ^a
\left( \psi_{-\beta}\left( \lambda^a \right) ^{\beta}_{\alpha}\right)
\nonumber\\ & -& \frac{i\nu}{3}\frac{d-1}{2C_{\rho}} \left(
\psi_{-\gamma}(\lambda^a)^{\gamma\beta}\psi_{-\beta} \right) \left(
\psi_{-\delta}(\lambda^a)^{\delta}_{\alpha} \right) ] \label{QQQ2}\\
\end{eqnarray}      
where the time derivatives are expressed in terms of the canonical
variables
using the equations of motion.
The Poisson brackets of the generators are
those of the superconformal algebra.  This can be directly verified
from (\ref{LLL}), (\ref{BBB}) and (\ref{QQQ}) using
the Poisson brackets for the super-Liouville field components and their
spacetime derivatives that follow from the equations of motion
and the basic canonical brackets, but the property actually
also holds if one regards the super-Liouville field as free.  This
fact is familiar from chiral quantization \cite{Lys}.

The realization of the Poisson bracket superconformal algebra in terms of
the
super-Liouville fields can also be used in order to construct
realizations for the quantum superconformal algebras. The classical
expressions (\ref{LLL}), (\ref{BBB}) and (\ref{QQQ})
become the quantum generators, some of the coefficients
getting corrections of order $\hbar$.
The expressions with the correct normalizations appear
in a paper by Bina and
G$\ddot{u}$naydin \cite{BG}.

\subsection{Symmetry transformations}

One can work out  the superconformal transformations
of the fields
by computing
their Poisson brackets with the generators.
For completeness,  we list the infinitesimal
symmetry transformations of all fields for
the tilded chirality superconformal generators:

1. Transformations induced by the improved Virasoro generator,
parametrized by $\epsilon = \epsilon(x^{-})$:

\begin{eqnarray}
\delta\varphi & = & \partial_{-}\varphi \epsilon
                 +\frac{1}{2} \partial_{-}\epsilon \\
\delta\psi_{+\alpha} &  = & \partial_{-}\psi_{+\alpha} \epsilon \\
\delta\psi_{-\alpha} &  = &
\partial_{-}\psi_{-\alpha} \epsilon
                 +\frac{1}{2} \psi_{-\alpha} \partial_{-} \epsilon \\
\delta u &  = & \partial_{-}u \epsilon \\
\delta [u^{-1}\partial_{-}u ]^a & = &
             \partial_{-} ( \epsilon [u^{-1}\partial_{-}u ]^a )
\end{eqnarray}

2. Transformations induced by the superconformal generators, parametrized
by $\zeta_{\alpha}=\zeta_{\alpha} (x^{-})$ :

\begin{eqnarray}
\delta\varphi & = & \frac{1}{2}
\left( \zeta_{\alpha}\eta^{\alpha\beta}\psi_{-\beta} \right) \\
\delta\psi_{+\beta} & = & -ie^{\varphi} \zeta_{\alpha}
[u^{-1}]^{\alpha}_{\beta} \\
\delta\psi_{-\alpha} & = & -\frac{i}{\nu} \left[ \partial_{-}\zeta_{\alpha}
+\partial_{-}\varphi\zeta_{\alpha}
-[u^{-1}\partial_{-}u]^a \zeta_{\beta}(\lambda^a)^{\beta}_{\alpha}\right]  
\nonumber \\
&-&\frac{d-1}{2C_{\rho}}
(\psi_{-\gamma}(\lambda^a)^{\gamma\beta}\zeta_{\beta})
\psi_{-\delta}(\lambda^a)^{\delta}_{\alpha}
-\frac{1}{2}(\zeta_{\gamma}\eta^{\gamma\beta}\psi_{-\beta}) \psi_{-\alpha}
\\
\delta u & = & -\frac{d-1}{2C_{\rho}}uT^a
(\psi_{-\alpha}(\lambda^a)^{\alpha\beta} \zeta_{\beta} ) \\
\delta [u^{-1} \partial_{-}u]^a & = & \frac{d-1}{2C_{\rho}}
\left[ -\partial_{-} (\psi_{-\alpha} (\lambda^a)^{\alpha\beta}
\zeta_{\beta})
+f^{abc} (\psi_{-\alpha}(\lambda^b)^{\alpha\beta}\zeta_{\beta})
          [u^{-1}\partial_{-}u]^c \right]
\end{eqnarray}
where we used the Jacobi identity (\ref{Jacobi}) for the $\lambda$
matrices.

3. Transformations induced by the Kac-Moody generators, parametrized by
$\omega^a = \omega^a (x^{-})$ :

\begin{eqnarray}
\delta\varphi & = & 0 \\
\delta\psi_{+\alpha} & = & 0 \\
\delta\psi_{-\alpha} & = & \omega^a \psi_{-\beta}(\lambda^a)^{\beta}_{\alpha}\\
\delta u & = & u \omega^{a}T^{a} \\
\delta [ u^{-1} \partial_{-} u ]^a & = &
\partial_{-}\omega^a
-f^{abc}\omega^b [ u^{-1} \partial_{-} u ]^c
\end{eqnarray}

One can easily verify that the super-Liouville action is indeed invariant
under
these transformations.

A special situation occurs for the $SU(1,1 \vert 2)/U(1)$
(``small $N=4$") case.  Writing the fermions as $2 \times 2$ matrices,
the super-Liouville action has an alternative $SU(2)$ symmetry obtained
by multiplication from the left of the fermionic matrix without involving
the WZ action.  This generates at the classical level an
alternative $N=4$ superconformal algebra.  The quantum realization
of this algebra was discussed  in \cite{KPR} and used recently
in a physically interesting context in
\cite{SW}.

\section{Classification of extended conformal algebras and spectral flow}
\setcounter{equation}{0}

Realizing the superconformal algebras as boundary theories of Chern-
Simons
actions allows a unified treatment of their classification following from
different possible ``moddings'' of the generators.
Some of these apparently
different algebras are related by the so called
``spectral flow `` transformations \cite{100}.
This feature also gets a general and natural
interpretation in the Chern-Simons framework .

We follow closely \cite{100}.
Both aspects mentioned above are related to the behaviour of the
fermionic generators under the extended Kac-Moody symmetry.
This behaviour, in turn, must be such that the action (\ref{ActionAction})
is single-valued.
For convenience, we will denote in this section:
$\psi_{\alpha}\equiv \psi_{+\alpha}$, $\zeta_{\alpha}\equiv
\psi_{-\alpha}$.

As discussed in the previous sections the physical Hilbert space is
realized
in terms of representations of the boundary superconformal algebra.
A given chiral representation is
defined  by  $h$~, the eigenvalue of the $L_0$ Virasoro operator and a
highest weight $\lambda$ of the Kac-Moody algebra.
Pairing the left and right representations, we construct the physical
Hilbert
space.
The highest weight $\vec{\lambda}$  is measured
by the holonomy $P exp (\int  i  B_{\phi}\,d\theta)$ around the boundary
of
the
disk.
For a given holonomy $exp(i 2 \pi \lambda) $, where
$\lambda \equiv \vec{\lambda}. \vec{H}$,  $H_i$ being the generators
in the Cartan subalgebra,
regularity at the center of the disk
(``origin")
requires the coupling
of the $B_0$ field to a source at the origin
in the representation $\lambda$  \cite{MSetAl}; the action
of the source is given by:
\begin{equation}
\int dt \,Tr [\lambda \omega^{-1}(\partial_0 +B_0)\omega (t)]
\label{2}
\end{equation}
Alternatively one can quantize the C-S theory on an
annulus  and require that
the holonomies on the two boundaries are the same.The source at the
center of the disk represents the zero- mode of the fields on the interior
boundary.

We start
by discussing the possible boundary conditions on the fields in
(\ref{ActionAction}). This will
provide us with the
classification of the superconformal algebras
living on the boundary.

The topology of the $AdS_3$ manifold is $D \times\Re$.
Parametrizing $D$ by $ r,\theta$ the fields could have  nontrivial
periodicity
conditions as  a function of $\theta$ provided the lagrangian density
remains
strictly periodic (single valued). In a real basis, the kinetic terms
of the $\psi_{\alpha}$, $\zeta_{\alpha}$ are invariant under a $O(d)$
transformation, where $d$ is the dimension of the real representation
$\rho$
of the compact  gauge group $\widetilde{G}$
(extended symmetry) under which the
gravitini transform.  In this section, $\widetilde{G}$ denotes the
internal group (and not its Lie algebra).
Therefore one can have the nontrivial periodicity
conditions:
\begin{equation}
\psi(\theta + 2\pi)=g \psi(\theta) \hspace{0.6 in}
\zeta(\theta + 2\pi)=\zeta(\theta) g^{-1}
\label{3}
\end{equation}
where $g$ is an arbitrary element of $O(d)$ and we made explicit just
the $\theta$ dependence of the fields.
We can make, however, a change of
variables in the functional integral
$\psi =\bar g \psi$ , $ \zeta=\zeta \bar g^{-1}$
which would leave the kinetic term form invariant. As a consequence
$g$ and $\bar g^{-1} g \bar g$ in (\ref{3}) are equivalent and the
boundary
conditions depend really only on conjugacy classes of $O(d)$. Therefore
the  $g$ in (\ref{3}) can be taken to be a generic element in the Cartan
torus
of $SO(d)$ or one  fixed matrix  $ g_0$ in $O(d)$
with determinant $-1$.

We start analyzing the continuous, Cartan torus part. In the presence of
the gauge
couplings,  $O(d)$ is not anymore a symmetry. The continuous part
is broken to $\widetilde{G} \times F $, $\widetilde{G}$
being the gauge group and $F$ a residual
``flavour'' group. For the supergroups listed in Table 1
the flavour group exists just for
  $ SU(1,1|2)/U(1)$
when it is
$SU(2)$.

It is convenient   to
consider the fermionic fields in the  Cartan basis
of $\widetilde{G}$, i.e.
$\psi_{\vec{\mu}}$ , $\zeta_{\vec{\mu}}$, where $\vec{\mu}$
is a weight
of the gauge group $\widetilde{G}$
in the representation $\rho $. In this basis
the transformations (\ref{3}) become phase transformations. In order to
keep the
lagrangian strictly periodic also the gauge fields $B$ should acquire
nontrivial periodicity conditions: the Kac-Moody algebra must be
``twisted"
\cite{GO}.

These are best studied if we use
the Cartan basis for the gauge fields,  i.e., we expand the connection $B$
in terms of the raising and lowering operators $E_{\vec{\alpha}}$ and the
operators in the Cartan subalgebra $H_i$
  $i=1,...,r$,
$\vec{ \alpha}$ being roots and $r$  the rank
of the group $\widetilde{G}$:
\begin{equation}
B=B_{\vec{\alpha}} E_{\vec{\alpha}} + B_i H_i
\label{4}
\end{equation}
Then it is easy to see that the periodicity conditions which will
leave the lagrangian single valued are:
\begin{equation}
\psi^a_{\vec{\mu}}(\theta+ 2\pi)=exp (i\vec{ a}. \vec{\mu}) 
 \psi^a_{\vec{\mu}}(\theta) \hspace{.6in}
  \zeta^a_{\vec{\mu}}(\theta+ 2\pi)=exp(i \vec{a}. \vec{\mu})
  \zeta^a_{\vec{\mu}} (\theta)
\label{5a}
\end{equation}
 and
\begin{equation}
B^a_{\vec{\alpha}} (\theta + 2\pi)= exp(i \vec{a}. \vec{\alpha} )
B^a_{\vec{\alpha}} (\theta) \hspace{.6 in}
B^a_i(\theta +2 \pi)=B^a_i(\theta)
\label{5b}
\end{equation}
where $ \vec{a}$ has $r$ continuous components.

In addition, for the $SU(1,1|2)/U(1)$
 case,
when a flavour symmetry exists, there is an overall $U(1)$
phase transformation (the Cartan torus of the flavour group)
affecting only the fermions:
\begin{equation}
\psi_{\vec{\mu}}(\theta + 2\pi)=exp (i b) \psi_{\vec{\mu}}(\theta)
\hspace{.6 in} \zeta_{\vec{\mu}} (\theta + 2 \pi)=exp (- i b)
\zeta_{\vec{\mu}}(\theta)
\label{6}
\end{equation}

We discuss now the periodicity conditions related to the discrete element
$g_0$ .
A necessary condition for the discrete operation $g_0$
to be compatible with the gauging is that each  gauge field acts in a real
representation. In particular, when the group $\widetilde{G}$ is not
simple,
each component of $\rho$ corresponding to a factor of
$\widetilde{G}$ should be real.

The action of the discrete element  $g_0$ on the fermions induces an
action on the gauge fields through the conjugation of the representation
matrices. When $ d$ is odd, $ g_0$ can be taken proportional to the
unit matrix and therefore commutes with the representation matrices  and
the gauge fields are not affected. This situation is realized for
$Osp(N|2;R)$ \ $N=odd$ and $G(3)$ in Table 1 which could have  therefore,
``discrete twisted'' boundary conditions for the fermions.
When $d$ is even the representation matrices are affected and the
transformation is compatible with gauging if it corresponds to an outer
automorphism of the group. This situation is realized only for
the $ Osp(N|2;R)$ \ $N=even$ case when the
$SO(N)$ group has a $Z(2)$ automorphism: parity. Therefore
among the superconformal algebras only $Osp(N|2;R)$ for all $N$,
and $G(3)$ will have
discrete twisted versions.

The number of continuous parameters characterizing the algebra
can be easily read off from Table 1 :

\noindent
$ [N /2] $ for $Osp(N|2;R)$, $4$ for
$F(4)$, $2$ for $G(3)$, $N$ for $SU(1,1|N)$, $2$ for $SU(1,1|2)/U(1)$,
$M+1$ for $Osp(4^{\ast}|2M) $ and $2$ for $D(2,1;\alpha)$.

However, algebras
having different values of the continuous parameters can be isomorphic.
This feature, the ``Spectral Flow'',
can be easily understood in the C-S framework.
Indeed, we can make a change of variable from the fields having nontrivial
periodicity properties (\ref{5a}), (\ref{5b}),  which have
an upper index $a $, to fields which are strictly
periodic and which we denote
without an index:
\begin{equation}
\psi^{a}_{\vec{\mu}}(\theta)= exp ( i \vec{a}.\vec{\mu}\, \theta /2 \pi)
\psi _{\vec{\mu}} (\theta) \hspace{.6 in}
B^a_{\vec{\alpha}} (\theta)= exp( i \vec{a}.\vec{\alpha}\, \theta /2 \pi)
B_{\vec{\alpha}}(\theta)
\label{7}
\end{equation}
We supplement (\ref{7}) with a change of variable also for the fields
$B^a_i$
which were periodic:
\begin{equation}
B^a_i(\theta)= B_i(\theta) + a_i/2\pi
\label{8}
\end{equation}

The change of variables (\ref{7}), (\ref{8}) corresponds formally to a
gauge
transformation:
\begin{equation}
u(\theta)= exp( i \vec{a}.\vec{H}\, \theta /2\pi)
\label{9}
\end{equation}
Since the gauge group element $u$ is not periodic, as a
gauge transformation (\ref{9}) is not legal.
Nevertheless we can use it as a change
of variable which will leave the bulk part , Eq.(\ref{ActionAction}),
of the C-S action
form invariant when expressed in terms of the unindexed, strictly periodic
variables.

The source term, Eq.(\ref{2}), will not be invariant. The change
can be easily calculated from the consistency requirement of the holonomy.
Since through the change of variables
an amount $ exp( i \vec{a}.\vec{H}/2 \pi ) $
was removed from the holonomy, when written in terms of the unindexed
variables the source term should have  $\lambda - a/2 \pi $
instead of $ \lambda $.

Therefore {\it all} dependence on continuous parameters in the boundary
conditions  which can be related to the gauge group
$\widetilde{G}$ is removable.
It is obvious that a similar procedure for the continuous dependence
related to the flavour symmetry or for the discrete boundary conditions
would not apply.

The above discussion can be used in a straightforward fashion
to obtain properties of the superconformal algebras. As discussed
in Section 3 the C-S fields become directly the generators of the
algebras.
Therefore the possible periodicity conditions of the generators
of the superconformal transformations $Q$ are identical to
the conditions (\ref{5a}) of the gravitino fields once a Cartan basis is
used.
The accompanying periodicity conditions of the Kac-Moody generators
$ B$ follow from the periodicity conditions of the gauge fields
(\ref{5b}).
The enumeration given above of the possible boundary conditions
for the various C-S theories applies
directly to the possible periodicities (``moddings'') of the generators
in the corresponding superconformal algebras.

 Thus the change of variable removing dependences on continuous
parameters gives the relation between algebras with different moddings,
i.e.
the spectral flow:
\begin{equation}
Q^a_{\vec {\mu}} (\theta) = exp(i \vec{a}. \vec{\mu}\, \theta/2 \pi)
Q_{\vec {\mu}} (\theta)
\label{10a}
\end{equation}
\begin{equation}
B^a_{\vec {\alpha}}(\theta) = exp ( i \vec{a}. \vec{\alpha}\, \theta/2
\pi)
B_{\vec {\alpha}} (\theta) \hspace{.6in}
B^a_i (\theta)= B_i(\theta) + a_i/2 \pi
\label{10b}
\end{equation}
where we used the Cartan basis for the generators.

The highest weight $\vec{ \lambda ^a}$ irrep of the algebra with
modding ``a'', is mapped into an irrep with highest weight $\vec{\lambda}$
of the  Ramond (completely periodic) algebra. The weights are related by:
\begin{equation}
\vec{\lambda^a}= \vec{\lambda} + \vec{a}/ 2 \pi
\label{11}
\end{equation}

We remark that since the spectral flow involves only the fields carrying
Kac-Moody quantum numbers, it wouldn't affect the part of the Virasoro
generator coming from the $sl(2)$ fields. The flow in the total
Virasoro generator can be read off from the flow of the Sugawara part
which was added to it, as discussed in Section 3.
{}From the explicit
Sugawara expression in terms of $B$ and  Eq. (\ref{10b}) we obtain
the flow of the Virasoro generator $L(\theta)$:
\begin{equation}
\widehat{L}^a (\theta)= L(\theta) + 1/[\pi(k+h)]\, \vec {B}(\theta). \vec{a} +
                 1/[4 \pi^2 (k+h)]\, \vec{a}.\vec {a}
\label{12}
\end{equation}

Summarizing:  besides
the completely periodic (Ramond) modding for all the cases listed in Table
1
the superconformal algebras admit the following {\it independent}
moddings:

 a) discrete twisted modding for the $Osp(N|2;R)$
 series (including the $N=2 ,3$
 and ``large'' $N=4$ superconformal algebras studied before ) and $G(3)$ .

 b) a  modding depending on one continuous parameter for the
 $SU(1,1|2)U(1)$ ( ``small'' $N=4$) superconformal algebra.

All the other moddings for all the algebras are related to the above
by spectral flow.

\section{Conclusions}
\setcounter{equation}{0}

In this paper, we have analyzed in detail the AdS/CFT correspondence for
extended
supergravity in $2+1$ dimensions.  We have shown how the superconformal
algebras with nonlinearities in the currents arise as asymptotic
symmetry algebras.  We have also shown that the dynamics of the asymptotic
fields
is controlled by the super-Liouville Lagrangian.

The AdS/CFT correspondence enlightens the properties of the non-linear
superconformal algebras and provides a systematic discussion of the
spectral
flow.

We conclude by pointing out two questions that could be worth
pursuing.
First, it would be interesting to try to understand the mechanism of
\cite{GS} from a pure three-dimensional point of view (see
\cite{STT} in this context).
Second, one might extend the analysis to the AdS$_2$/Super-conformal
quantum mechanics along the lines of \cite{MS1,MS2,HP}.
The extensions to the quasiconformal case (in the sense of \cite{Frad3})
and to Toda models (super $W$-algebras)
\cite{PhysRep} could also be of interest.

\section*{Acknowledgements}

We are grateful to Karin Bautier for discussions on the
earlier parts of this article.
This work has been partly supported by the ``Actions de
Recherche Concert{\'e}es" of the ``Direction de la Recherche
Scientifique - Communaut{\'e} Fran{\c c}aise de Belgique", by
IISN - Belgium (convention 4.4505.86) by
Proyectos FONDECYT 1970151 and 7960001 (Chile), by
Center for Basic Research of the Israeli Academy of Sciences
and by the Minerva Foundation, Germany.

\section*{Appendix}
\renewcommand{\theequation}{A.\arabic{equation}}
\setcounter{equation}{0}

In this appendix, we indicate where zero modes and holonomies
arise in the derivation of the super-Liouville action
from the supergravity action.  We show in particular that
2+1 supergravity only yields the ordinary super-Liouville
Lagrangian modulo zero mode and holonomy terms, one striking
difference being that the holonomies in the right and left
sectors are independent in supergravity.
The analysis sheds as a by-product some interesting
light on the sign of the exponential term in the Liouville
action.

Since the subtleties arise already in the non-supersymmetric case,
we shall drop the fermions in order to keep the formulas
simple.  Thus, our starting point is the 
action for ordinary gravity, 
\begin{equation}
S\left[A,\widetilde{A}\right] = \frac{1}{8\pi G}
{\int}_{M}d^{3}x\{\frac{1}{2}eR+\frac{e}{\ell^{2}} \}
\end{equation}
where $A$ and $\widetilde{A}$ are related to the dreibeins and the 
connection through (\ref{dreib}) and (\ref{dreibein}).
This action can be written as the difference between two
$sl(2,R)$-Chern-Simons actions.  

The first step follows the references
\cite{MS,MSetAl,CHvD} and consists
in solving the Gauss constraints $F^a_{ij} = 0$, $\widetilde{F}^a_{ij}
=0$ ($i,j$ spatial indices) in the bulk,
taking into account the first set of boundary conditions
(\ref{bc1}).
We shall assume that the spatial sections have the annulus 
topology, as this is the case relevant to the black holes.
As shown in \cite{MSetAl}, the solution of the Gauss constraints
is then
\begin{eqnarray}
A_\theta &=& g^{-1} K(t) g + g^{-1} \partial_\theta g, \\
A_r &=& g^{-1} \partial_r g 
\end{eqnarray}
where $g(t,r, \theta)$ is a $sl(2,R)$ group element and where
$K(t)$ is a Lie-algebra element characterizing the holonomy.
Under the transformation
$g(t,r, \theta) \rightarrow \alpha(t) g(t,r, \theta)$,
$K(t)$ is conjugated by $\alpha(t)$.  Using this freedom,
one can bring $K(t)$ along $\sigma^3$ (the black hole
holonomy is hyperbolic and we restrict therefore
the analysis to this case),
\begin{equation}
K(t) = K^3(t) \frac{\sigma^3}{2}.
\end{equation}
Similar formulas hold for the tilde sector.

If one inserts the solution of the Gauss constraints inside
the Chern-Simons action supplemented by the surface term
appropriate to the boundary conditions
(\ref{bc1}), one gets
\begin{equation} 
S = S^\infty + S^{Hor}
\end{equation}
where $\infty$ and $Hor$ refer respectively to the outer and
inner boundaries.  We are interested here in the piece
$S^\infty$ describing the dynamics at infinity,
given by
\begin{equation}
S^\infty = I + \widetilde{I}
\end{equation}
with
\begin{eqnarray}
I[h(t,\theta), K(t)] &=& \frac{k}{4 \pi} \int dt d \theta
\left[ Tr\left( h^{-1} \partial_\theta h h^{-1} \partial_t h
\right) - Tr E^2_\theta \right]
\nonumber \\
& & +  \frac{k}{2 \pi} \int dt d \theta \left[ 
Tr \left( h^{-1} K(t) 
\partial_t h \right) \right] \nonumber \\
& & +  \frac{k}{4 \pi} \int dr dt d\theta Tr
\left[ g^{-1} \partial_r g \left( g^{-1} \partial_t g
g^{-1} \partial_\theta g - g^{-1} \partial_\theta g 
g^{-1} \partial_t g \right) \right].
\end{eqnarray}
Here, $h(t,\theta)$ is the value of $g(t,r, \theta)$ on the outer boundary
(we assume that the $r$-dependent similarity transformation of subsection
3.2 has been performed, so that the fields have well-defined asymptotics)
and 
\begin{equation}
E_\theta = h^{-1} K(t) h + h^{-1} \partial_\theta h
\end{equation}
The holonomy $K(t)$ appears also in the action $S^{Hor}$ describing
the dynamics at the inner boundary, but not $h(t, \theta)$. 
Similarly, the action for the other chirality is
\begin{eqnarray}
\widetilde{I}[\widetilde{h}(t,\theta), \widetilde{K}(t)] &=&
\frac{k}{4 \pi} \int dt d \theta
\left[ Tr\left(- \widetilde{h}^{-1} \partial_\theta \widetilde{h}
\widetilde{h}^{-1} \partial_t \widetilde{h} \right) -
Tr \widetilde{E}^2_\theta \right]
\nonumber \\
& & -  \frac{k}{2 \pi} \int dt d \theta \left[ Tr
\left( \widetilde{h}^{-1} \widetilde{K}(t) \partial_t \widetilde{h}
\right) \right] \nonumber \\
& & -  \frac{k}{4 \pi} \int dr dt d\theta Tr
\left[\widetilde{g}^{-1} \partial_r \widetilde{g}\left(
\widetilde{g}^{-1} \partial_t \widetilde{g}
\widetilde{g}^{-1} \partial_\theta \widetilde{g} -
\widetilde{g}^{-1} \partial_\theta \widetilde{g}
\widetilde{g}^{-1} \partial_t \widetilde{g}
\right) \right]
\end{eqnarray} 
with 
\begin{equation}
\widetilde{E}_\theta = \widetilde{h}^{-1} \widetilde{K}(t)
\widetilde{h} + \widetilde{h}^{-1} \partial_\theta \widetilde{h}
\end{equation}
and
\begin{equation}
\widetilde{K}(t) = \widetilde{K}^3(t) \frac{\sigma^3}{2}
\end{equation}

At this point, one can proceed in two different ways.  One can either
first recombine the two chiralities to get the vector WZW model
(modulo holonomies and zero mode terms) 
and then, implement the Hamiltonian reduction enforced by the
second set of constraints (\ref{bc2}).  This was the method
followed  in  \cite{CHvD}.  Alternatively,
one can first implement the
reduction constraints (\ref{bc2}) and then discuss how to recombine the
two (reduced) chirality sectors.
This is the approach followed here.

Consider first the untilde sector.  Using the Gauss decomposition\footnote{The
global aspects of the Gauss decomposition are discussed in \cite{GGG}.}
\begin{equation}
h = \exp{(\alpha \sigma^-)} \exp{(\beta \sigma^3)}
\exp{(\gamma \sigma^+)}
\end{equation} 
one finds that the action takes the form
\begin{equation}
I = \frac{k}{\pi}\int dt d\theta \left[ \partial_- \beta \partial_\theta \beta
+ \partial_\theta \alpha \partial_- \gamma  \exp{(2 \beta)} \right]
+ \frac{k}{\pi} \int dt d\theta Tr\left(h^{-1} K(t) \partial_- h \right)
- \frac{k}{4} \int dt (K^3)^2
\end{equation}
with $Tr\left(h^{-1} K(t) \partial_- h \right)$ given explicitly by
\begin{equation}
Tr\left(h^{-1} K(t) \partial_- h \right) = 
K^3(t)\left[ \partial_{-}\beta -\alpha \partial_{-}\gamma exp(2\beta)
\right]
\end{equation}
while the constraint (\ref{bc2}) on the untilde current reads
\begin{eqnarray}
\partial_\theta \alpha \exp{(2 \beta)} - \alpha K^3 \exp{(2 \beta)}  &=& \mu.
\\
\partial_\theta \beta - \gamma \partial_\theta \alpha
\exp{(2 \beta)} + \frac{K^3}{2} + \alpha \gamma K^3 \exp{(2 \beta)} &=& 0
\end{eqnarray}
The first of these constraints enables one to express $\alpha$ in terms
of $K^3$ and $\beta$; this can be verified by Fourier expansion of $\alpha$
and $\exp{(- 2 \beta)}$; note the crucial role played by the holonomy, which
must be non zero for the zero-mode part of the equation
to make sense.  The relationship between $\alpha$ and $\beta$ is non-local
in $\theta$.  The second of these constraints can be rewritten as
\begin{equation}
\partial_\theta \beta - \mu \gamma + \frac{K^3}{2} = 0
\end{equation}
and enables one to express $\gamma$ in terms of $K^3$ and $\beta$
($\mu \not= 0$).

Thus, one can completely get rid of $\alpha$ and $\beta$.  If one does so, 
one gets a reduced action $I^R[\beta(t, \theta), K^3(t)]$ that involves
only $\beta(t, \theta)$ and $K^3(t)$.  Its  expression is 
straightforward to work out but it is rather awkward
and will not be given here. 
Suffice it to note that the
$K^3$-independent part of $I^R$ is just the action
for a free chiral boson \cite{JF}
\begin{equation}
I^R = \frac{k}{\pi}\int dt d\theta \left[ \partial_- \beta \partial_\theta \beta
\right] + (K^3-terms)
\label{Ireduced}
\end{equation}
and does not contain $\mu$ (in particular, it is independent
of the sign of $\mu$).  The (unwritten) piece involving $K^3$ in (\ref{Ireduced})
is anyway incomplete since $K^3$ occurs also in the action $S^{Hor}$
describing the dynamics on the inner boundary.  
Similarly, the improved Virasoro generator is given by
\begin{equation}
\widehat{L} = - \frac{k}{2 \pi}
\left[(\partial_\theta \beta)^2 + \partial_\theta \partial_\theta \beta
\right]
\end{equation}
(the minus sign is due to our conventions) and again, does not depend on $\mu$.

The same treatment can be worked out for the other chirality, yielding
\begin{equation}
\widetilde{I}^R = -\frac{k}{\pi}\int dt d\theta \left[ \partial_+ 
\widetilde{\beta} \partial_\theta \widetilde{\beta}
\right] + (\widetilde{K}^3-terms)
\end{equation}
and
\begin{equation}
\widehat{\widetilde{L}} = \frac{k}{2 \pi}
\left[(\partial_\theta \widetilde{\beta})^2 - \partial_\theta \partial_\theta
\widetilde{\beta}\right].
\end{equation}
The first term is again the action for a chiral boson, of opposite
chirality.
Note that the Hamiltonians for the chiral bosons (dropping
holonomy terms) are both positive
definite.

One can now recombine the two chiralities.  
First, consider the solutions of the equations of motion.
The chiral fields $\beta$ and $\widetilde{\beta}$
depend on $x^+$ and $x^-$, respectively.
Define
\begin{equation}
\exp{2\varphi} = \frac{\partial_+ F 
\partial_- \widetilde{F}}{[1 - F \widetilde{F}]^2}
\label{defini}
\end{equation}
with $\partial_+F= \exp{(-2\beta)}$ and $
\partial_- \widetilde{F}= \exp{(2\widetilde{\beta})}$.
It is easy to see that $\varphi$ is a solution of the Liouville
equation
\begin{equation}
\partial_+ \partial_- \varphi - \exp{2\varphi} = 0.
\end{equation}
If one makes the corresponding formal change of phase space variables
\begin{eqnarray}
\varphi &=& - \beta + \widetilde{\beta} - \log |1- F \widetilde{F}|,
\; \; F' = \exp{(-2 \beta)}, \; 
\widetilde{F}' = \exp{2 \widetilde{\beta}} \\
\Pi_\varphi &=& \frac{k}{2 \pi} \left( - \beta' -\widetilde{\beta}'
+ \frac{\exp{(-2 \beta)} \widetilde{F} - F \exp{2 \widetilde{\beta}}}
{1- F \widetilde{F}} \right)
\end{eqnarray}
one gets from the sum of the chiral actions
for $(\beta, \widetilde{\beta})$ the Liouville action 
(\ref{superLiouvilleAction}) (with fermions and gauge fields dropped).
Of course, the above transformation must be amended so
as to be well-defined  on the circle; this leads to additional terms in the
super-Liouville action involving zero modes and holonomies,
which are crucial (furthermore, we have not investigated the
global features of the change of variables). 
Thus, the action coming
from 2+1 gravity is not just the vector Liouville action, 
there are further zero modes and holonomy terms.
In particular, the 
holonomies $K^3$ and $\widetilde{K}^3$ are
independent in 2+1 gravity.
If the difference of the holonomies is an integer (i.e. the zero-mode
conjugate to the difference is an angular momentum variable) the CFT on
the boundary will still be local.
We shall not discuss explicitly the extra terms here, because
for practical purposes, it is more convenient not to recombine
the chiralities and to regard
the sum of the chiral actions (with the holonomy terms)
as defining the Liouville model relevant to the asymptotic of $2+1$ 
gravity\footnote{Thus, in particular, the super-Liouville action
given in the text does not fully describe 2+1 supergravity. The
zero modes and holonomies are different.  Nevertheless, deriving
the super-Liouville action from 2+1 supergravity (up to
zero modes) is useful since it enables one to get straightforwardly
the superconformal transformation laws.}.
In the case of supergravity models 
with different number of supersymetries in the right and left sectors, this
is the only approach possible. 

Note that alternatively, we could have defined
\begin{equation}
\exp{2\Phi} = \frac{\partial_+ F
\partial_- \widetilde{F}}{[1 + F \widetilde{F}]^2}
\end{equation}
The variable $\Phi$ is a solution of the Liouville equation with 
standard sign,
\begin{equation}
\partial_+ \partial_- \Phi + \exp{2\Phi} = 0.
\end{equation}
The corresponding phase space change of variables, familiar from
the analysis of the standard Liouville model \cite{GN,DHJa,BCT}
\begin{eqnarray}
\Phi &=& - \beta + \widetilde{\beta} - \log |1+ F \widetilde{F}|,
\; \; F' = \exp{(-2 \beta)}, \;
\widetilde{F}' = \exp{2 \widetilde{\beta}} \\
\Pi_\Phi &=& \frac{k}{2 \pi} \left( - \beta' -\widetilde{\beta}'
- \frac{\exp{(-2 \beta)} \widetilde{F} - F \exp{2 \widetilde{\beta}}}
{1+ F \widetilde{F}} \right)
\end{eqnarray}
brings the action to the standard Liouville action.  Of course, the zero modes
and holonomies enter differently this time.  But, up to these terms, we see
that one can change the sign in front of the Liouville exponential
in the action by a (non-local, real) change of variables.


\begin{thebibliography}{99}
\bibitem{Maldacena} J. Maldacena, {\it The Large N Limit of Superconformal
Field
Theories and Supergravity}, Adv. Theor. Math. Phys.
{\bf 2} (1998) 231, hep-th/9711200.
\bibitem{GP} S.S Gubser, I.R. Klebanov and A.M. Polyakov,
{\it Gauge theory correlators from noncritical string theory},
Phys. Lett. {\bf B 428} (1998) 105, hep-th/9802109.
\bibitem{Witten} E. Witten, {\it Anti-de Sitter space and holography},
Adv. Theor. Math. Phys.
{\bf 2} (1998) 253, hep-th/9802150.
\bibitem{review} O. Aharony, S. S. Gubser, J. Maldacena, H. Ooguri and Y.
Oz,
{\it Large $N$ field theories, string theory and gravity},
hep-th/9905111.
\bibitem{DJT} S. Deser, R. Jackiw and S. Templeton, 
{\it Three-Dimensional Massive Gauge Theories}, Phys. Rev. Lett.
{\bf 48} (1982) 975.
\bibitem{W0} E.Witten, {\it Non-Abelian Bosonization in Two Dimensions},
Commun. Math. Phys. {\bf 92} (1984), 455-472.
\bibitem{W1} E. Witten, {\it Quantum Field Theory and the Jones
Polynomial}
Commun. Math. Phys. {\bf 121} (1989) 351.
\bibitem{MS} G. Moore and N. Seiberg, 
{\it Taming the Conformal Zoo}, Phys. Lett. {\bf 220 B}
(1989) 422.
\bibitem{MSetAl} S.Elitzur, G.Moore, A. Schwimmer, N.Seiberg,
{\it Remarks on the Canonical Quantization of the Chern-Simons-Witten
Theory,}
Nuc.Phys. {\bf B326}(1989)108-134.
\bibitem{AT} A.Achucarro and P.K.Townsend, {\it A Chern-Simons Action for
Three-Dimensional Anti-de-Sitter Supergravity Theories}, Phys.Lett.
{\bf B180} (1986) 89-92.
\bibitem{W2} E. Witten, {\it 2+1 Dimensional Gravity as an Exactly
Solvable  System}, Nucl. Phys {\bf B311} (1988/89) 46-78.
\bibitem{CHvD} O. Coussaert, M. Henneaux and P. van Driel,
{\it The asymptotic
dynamics of three-dimensional Einstein gravity with a negative
cosmological
constant}, Class. Quant. Grav. {\bf 12} (1995) 2961, gr-qc/9506019 v2.
\bibitem{BH} J.D. Brown and M.Henneaux, {\it
Central Charges in the
Canonical Realization of Asymptotic Symmetries: An Example from Three
Dimensional Gravity}, Commun.Math.Phys. {\bf 104} (1986) 207-226.
\bibitem{X} A. Alekseev and S. Shatashvili, 
{\it Path Integral Quantization of the Coadjoint Orbits of the
Virasoro Group and 2-D Gravity}, Nucl. Phys.
{\bf B323} (1989) 719.
\bibitem{XX} M. Bershadsky and H. Ooguri, 
{\it Hidden SL(N) Symmetry in Conformal Field Theories},
Commun. Math. Phys.
{\bf 126} (1989) 49.
\bibitem{XXX} P.
Forg\'acs, A. Wipf, J. Balog, L. Feh\'er and L. O'Raifeartaigh,
{\it Liouville and Toda Theories as Conformally Reduced WZNW
Theories}, 
Phys. Lett. {\bf 227 B} (1989) 214.
\bibitem{XXXX} L. O'Raifeartaigh and V.V. Sreedhar,
{\it Path integral formulation of the conformal
Wess-Zumino
-Witten $\rightarrow$ Liouville reduction},
Phys. Lett. {\bf B425} (1998) 291.
\bibitem{CH} O. Coussaert and M. Henneaux, {\it
Supersymmetry of the 2+1 black hole}, hep-th/9310194,
Phys. Rev. Lett. {\bf 72} (1994) 183.
\bibitem{BBCHO} M. Banados, K. Bautier, O. Coussaert, M. Henneaux
and M. Ortiz, {\it Anti-de Sitter/CFT correspondence in three-dimensional
supergravity}, Phys. Rev. {\bf D58} (1998) 085020, hep-th/9805165;
see also K. Bautier, {\it AdS$_3$ Asymptotic (Super)symmetries}, 
hep-th/9909097.
\bibitem{Y} V. G. Knizhnik, 
{\it Superconformal Algebras in Two-Dimensions},
Theor. Math. Phys. {\bf 66} (1986) 68.
\bibitem{YY} M.Bershadsky, {\it Superconformal Algebras in Two Dimensions
with arbitrary N}, Phys.Lett. {\bf B174}(1986) 285-288.
\bibitem{Frad1} E.S. Fradkin and V. Ya Linetsky, {\it An
exceptional $N=8$ superconformal algebra in two dimensions
associated with $F(4)$}, Phys. Lett. {\bf B275} (1992) 345.
\bibitem{Frad2} E.S. Fradkin and V. Ya Linetsky, {\it Results
of the classification of superconformal algebras in two dimensions},
Phys. Lett. {\bf B282} (1992) 352.
\bibitem{Frad3} E.S. Fradkin and V. Ya Linetsky, {\it Classification of
superconformal and quasiconformal algebras in two dimensions},
Phys. Lett. {\bf B291} (1992) 71.
\bibitem{Bowcock} P.Bowcock, {\it Exceptional Superconformal Algebras}
Nucl. Phys. {\bf B381} (1992) 415.
\bibitem{100}
A. Schwimmer and N.Seiberg, 
{\it Comments on the N=2,N=3,N=4 Superconformal Algebras in
Two-Dimensions}, Phys. Lett. {\bf B184}
(1987) 191 .
\bibitem{deBoer} Jan de Boer, {\it Six-Dimensional Supergravity on
$S^{3}\times AdS_{3}$ and 2d Conformal Field Theory},
Nucl. Phys. {\bf B548} (1999) 139, hep-th/9806104v2.
\bibitem{Ishimoto} Y. Ishimoto, {\it Classical Hamiltonian Reductioon
on $D(2\vert 1; \alpha)$ Chern-Simons theory and large $N=4$
superconformal
theory}, hep-th/9808094.
\bibitem{Ito} K. Ito, {\it Extended superconformal algebras on $AdS_3$},
hep-th/9811002,
\bibitem{David} J.R. David, G. Mandal, S. Vaidya and S. R. Wadia,
{\it Point mass geometries, spectral flow and AdS$_3$-CFT$_2$
correspondence},
hep-th/9906112.
\bibitem{Kac} V.G. Kac, {\it A sketch of Lie Superalgebra Theory},
Commun. Math. Phys {\bf 53} (1977) 31; {\it Lie Superalgebras}, 
Adv. Math. {\bf 26} (1977) 8; Funct. Anal. {\bf 9} (1975) 91.
\bibitem{Nahm} W.Nahm, {\it Supersymmetries and their Representations},
Nucl. Phys. {\bf B135} (1978) 149-166.
\bibitem{Gun} M.G\"{u}naydin, G.Sierra and P.K.Townsend, {\it The Unitary
Supermultiplets of d=3 AdS and d=2 Conformal Superalgebras}, Nucl. Phys.
{\bf B264} (1986) 429.
\bibitem{BG} B. Bina and M.G\"{u}naydin, {\it Real Forms of
Nonlinear Superconformal and Quasi-Superconformal Algebras and Their
Unified Realization}, Nucl. Phys. {\bf B502} (1997) 713.
\bibitem{DJ} S. Deser and R. Jackiw, {\it Three-dimensional cosmological
gravity: dynamics of constant curvature}, Ann. Phys. (N.Y.)
{\bf 153} (1984) 405.
\bibitem{BTZ} M.Ba\~{n}ados, C.Teitelboim, J.Zanelli, {\it The Black Hole
in
Three Dimensional Space Time}, Phys. Rev. Lett. {\bf 69} (1992) 1849,
hep-th/9204099v2.
\bibitem{BTZH} M.Ba\~{n}ados, M.Henneaux, C.Teitelboim, J.Zanelli,
{\it Geometry of the 2+1 black hole}, Phys.Rev.D. {\bf 48}
(1993)1506-1525.
\bibitem{Teit} M.Henneaux and C.Teitelboim, {\it Asymptotically Anti-de
Sitter Spaces}, Commun.Math.Phys {\bf 98}(1985), 391-424.
\bibitem{RT} T. Regge and C. Teitelboim, 
{\it Role of Surface Integrals in the Hamiltonian Formulation o
General Relativity},
Ann. Phys. (N.Y.)
{\bf 88} (1974) 286.
\bibitem{NN} J. Navarro-Salas and P. Navarro, {\it A note on Einstein
gravity on $AdS_3$ and boundary conformal field theory},
Phys. Lett. {\bf 439} (1998) 262, hep-th/9807019.
\bibitem{Ba} M. Ba\~nados and M.E. Ortiz, {\it The central charge in three
dimensional anti-de Sitter space}, Class. Quant. Grav. {\bf 16}
(1999) 1733, hep-th/9806089.
\bibitem{PW} A.M.Polyakov, P.B.Wiegmann, {\it Theory of Non-Abelian
Goldstone
Bosons}, Phys.Lett. {\bf 131B} (1983) 121.
\bibitem{Lys} L.Caneschi, M.Lysiansky,
{\it Chiral Quantization of the WZW SU(n) Model},
Nucl. Phys. {\bf B505} (1997) 701, hep-th/9605099.
\bibitem{HT} M. Henneaux and C. Teitelboim, Quantization of gauge systems,
Princeton University Press, Princeton: 1992.
\bibitem{BO} M. Bershadsky and H. Ooguri, 
{\it Hidden OSP(N,2) Symmetries in Superconformal Field Theories},
{\em Phys. Lett.} {\bf 229 B}
(1989) 374.
\bibitem{Z} A. Sevrin, K. Thielemans and W. Troost, {\it Induced and
Effective
Gravity Theories in $D=2$}, Nucl. Phys. {\bf B407} (1993) 459,
hep-th/9303133.
\bibitem{ZZ} K. Ito, J.O. Madsen and J.L. Petersen, {\it
Extended Superconformal Algebras from Classical and Quantum Hamiltonian
Reduction}, hep-th/921019.
\bibitem{KPR} C. Kounnas, M. Porrati and B. Rostand, {\it On $N=4$
Extended
 Super-Liouville Theory}, {\em Phys.Lett.} {\bf 258 B} (1991) 61.
\bibitem{ZH}Yao-Zhong Zhang, {\it $N$ Extended Super-Liouville Theory
 from $OSP(N/2)$ WZNW Model}, {\em Phys.Lett.} {\bf 283 B} (1992) 237.
\bibitem{SW} N. Seiberg and E. Witten, {\it The D1/D5 System and Singular
CFT} , hep-th/9903224
\bibitem{GO} P. Goddard and D. Olive, 
{\it Kac-Moody and Virasoro Algebras in Relation to Quantum
Physics},
Int. J.
Mod. Phys. {\bf A1} (1986) 303.
\bibitem{GS} P. Goddard and A. Schwimmer, {\it Factoring out free
fermions and superconformal algebras}, Phys. Lett. {\bf B214}
(1988) 209.
\bibitem{STT} A. Sevrin, K. Thielemans and W. Troost,
{\it The relation between linear and non-linear $N=3,4$ supergravity
theories}, Phys. Rev. {\bf D48} (1993) 2789, hep-th/9304020.
\bibitem{MS1} J. Michelson and A. Strominger,
{\it The geometry of (super) conformal quantum mechanics},
hep-th/9907191.
\bibitem{MS2} J. Michelson and A. Strominger, 
{\it Superconformal Multi-Blackhole Quantum Mechanics},
hep-th/9908044.
\bibitem{HP} S. Hellerman and J. Polchinski, {\it Supersymmetric Quantum
Mechanics from Light Cone Quantization}, hep-th/9908202.
\bibitem{PhysRep} L. Feh\'er, L.O'Raifeartaigh, P. Ruelle, I. Tsutsui and
A. Wipf, {\it On Hamiltonian Reductions of the WZNW theories},
Phys. Rep. {\bf 222} (1992) 1.
\bibitem{GGG} I. Tsutsui and L. Feher,
{\it Global aspects of the WZNW reduction to Toda
theories}, Progr. Theor. Phys. Suppl. {\bf 118} (1995) 173,
hep-th/9408065. 
\bibitem{JF} R. Floreanini and R. Jackiw, {\it Selfdual
fields as charge density solitons}, Phys. Rev. Lett. {\bf 59} (1987) 1873. 
\bibitem{GN} J.-L. Gervais and A. Neveu, {\it Dual string spectrum
in Polyakov's quantization. (II) Mode separation}, Nucl. Phys. {\bf B209}
(1982) 125.
\bibitem{DHJa} E. D'Hoker and R. Jackiw, {\it Classical and quantal
Liouville field theory}, Phys. Rev. {\bf D 26} (1982) 3517.
\bibitem{BCT} E. Braaten, T. Curtright and C. Thorn, {\it Quantum
B\"acklund transformation for the Liouville theory}, Phys. Lett. {\bf 118B}
(1982) 115. 

\end{thebibliography}
\end{document}